\documentclass{aastex631}

\usepackage[version=4]{mhchem}

\begin{document}
\title{Day-night transport induced chemistry and clouds on WASP-39b: Gas-phase composition}

\author[0000-0002-8163-4608]{Shang-Min Tsai}
\affiliation{Department of Earth and Planetary Sciences, University of California, Riverside, CA, USA}
\author[0000-0002-8837-0035]{Julianne I. Moses}
\affiliation{Space Science Institute, Boulder, CO, USA}
\author[0000-0002-4250-0957]{Diana Powell}
\affiliation{Center for Astrophysics ${\rm \mid}$ Harvard {\rm \&} Smithsonian, Cambridge, USA}
\affiliation{Department of Astronomy \& Astrophysics, University of Chicago, Chicago, IL, USA}
\author[0000-0002-3052-7116]{Elspeth K.H. Lee}
\affiliation{Center for Space and Habitability, University of Bern, Gesellschaftsstrasse 6, CH-3012 Bern, Switzerland}



\begin{abstract}
JWST has recently detected the first robust photochemical product on an exoplanet: sulfur dioxide (\ce{SO2}) on WASP-39b \citep{Zafra2023,Alderson2023,Tsai2023b}. The data from the NIRISS instrument also reveal signs of partial coverage of clouds \citep{Feinstein2023}. Most of the previous studies have focused on interpreting spectral data with 1D models. To explore how the chemical species and cloud particles are altered by global circulation, we applied a 2D photochemical model and a 2D microphysical cloud model separately to post-process the thermal and dynamical structures simulated by a 3D general circulation model (GCM) of WASP-39b. We found that \ce{SO2} produced by photochemistry on the dayside can be transported to the nightside owing to the efficient replenishment of horizontal transport. The morning-evening limb differences in methane (\ce{CH4}) abundances predicted by the 1D models disappeared after horizontal transport is included. Similarly, the inclusion of horizontal transport also reduced the limb differences in \ce{SO2}. Our modeling results suggest that the fast zonal wind results in minimal or negligible limb asymmetry in composition. Based on the synthetic spectra generated by our 2D atmosphere simulations, we propose that observing \ce{SO2} absorption in the emission spectra of WASP-39b at different phases may offer opportunities to probe the horizontal quenching process of photochemical products. We will focus on the gas-phase chemistry in this paper and leave the results regarding clouds in the subsequent paper as part of the series.



\end{abstract}


\section{Introduction}\label{sec:intro}
Recent JWST observations of WASP-39b have unveiled exoplanet atmospheric characteristics with unprecedented details \citep{ERS2023,Zafra2023,Alderson2023,Ahrer2023}. The first robust photochemical evidence via detecting \ce{SO2} in the \ce{H2}-dominated environment of WASP-39b marks a breakthrough in exoplanet atmospheric characterization \citep{Tsai2023b}. The derived bulk elemental constraints with the addition of sulfur are crucial to furthering our understanding of the formation and evolution of gas giants \citep[e.g.,][]{Turrini2021,Crossfield2023}. While we are still in the dawn of the JWST era, its ongoing programs and other upcoming observational facilities (e.g., ARIEL\footnote{The Atmospheric Remote-sensing Infrared Exoplanet Large-survey} and ELT\footnote{The Extremely Large Telescope}) set the stage for exquisite spectral data of gas giants. Comprehensive theoretical models are desired to interpret the rich dataset and reveal key insights into planetary properties.
 
Transmission spectroscopy plays a central role in the study of exoplanet atmospheric composition. The transmission spectra sample regions around the terminator and probe two opposite sides of the planet, during dawn and dusk. An increasing body of research has emphasized the significant influence of spatial inhomogeneities when performing atmospheric retrievals \citep{Line2016,MacDonald2020} and cross-correlation analysis on high-resolution data \citep{Joost2021}. \cite{Espinoza2021} and \cite{Grant2023} further presented methods to disentangle the components from the morning (leading) and evening (trailing) limbs. These techniques shed light on the driving force behind climate patterns on distant worlds, such as the equatorial jet \citep{Ian2012} and the redistribution of dayside-originated photochemical hazes and nightside-formed condensation clouds \citep{Kempton2017}. 

For WASP-39b, the compositional profiles of the two limbs have been simulated individually in \cite{Tsai2023b} with a suite of 1D photochemical models. These models anticipate notable differences in composition between the two limbs of WASP-39 b, given that 3D GCMs predict that the morning terminator would be much cooler than the evening terminator \citep{Venot2020,Baeyens2021}. However, the 1D photochemical models, while successful in explaining the observed spectra and the origin of \ce{SO2}, do not account for the impact of horizontal winds. Considering the relative inactivity of the host star \citep{Ahrer2023}, any observed limb asymmetries of WASP-39b could be attributed to planetary aspects rather than stellar effects (e.g., stellar spots, gravity darkening). With the extensive spectral data and chemical signatures obtained through transit observations, WASP-39b presents a promising opportunity to explore the global map of chemical composition and clouds. 





Considerable efforts have been made to incorporate cloud and haze across various levels of complexity in 3D GCMs \citep{Lee2016,Lines2018,Steinrueck2021}. The cutting-edge 3D GCMs also have the capability to include thermochemical reactions and account for transport-induced quenching, i.e. the homogenization of chemical composition caused by atmospheric dynamics \citep{Cooper2006,Mendonca2018a,Drummond2020,Zamyatina2023,Lee2023}. However, despite the ongoing efforts of applying Earth climate models \citep{Yates2020,Chen2021} or new approaches to simplify the chemical module in a 3D GCM \citep{Tsai2021b}, these advances have not yet been realized for gas giant atmospheres due to computational limitations. Pseudo-2D photochemical models \citep{Agundez2014,Venot2020b,Baeyens2022,Moses2021} that employ a rotating 1D column to mimic a uniform flow in the Lagrangian view have provided complementary tools to the traditional 1D photochemical-kinetics model and the 3D circulation model that does not treat chemical processes.  

In this work, following up on the identification of sulfur photochemistry, we investigate the influence of atmospheric circulation on the redistribution of composition in WASP-39b. We specifically examine the behavior of photochemically produced \ce{SO2} and explore the implications for transmission and emission spectral observations. By comparing our findings with traditional 1D models, we assess the validity and limits of the 1D approach. We will limit our discussion in this paper to the gas-phase chemical distribution, while the results concerning clouds will be presented in the subsequent paper.


\begin{figure}[ht!]
\includegraphics[width=0.495\columnwidth]{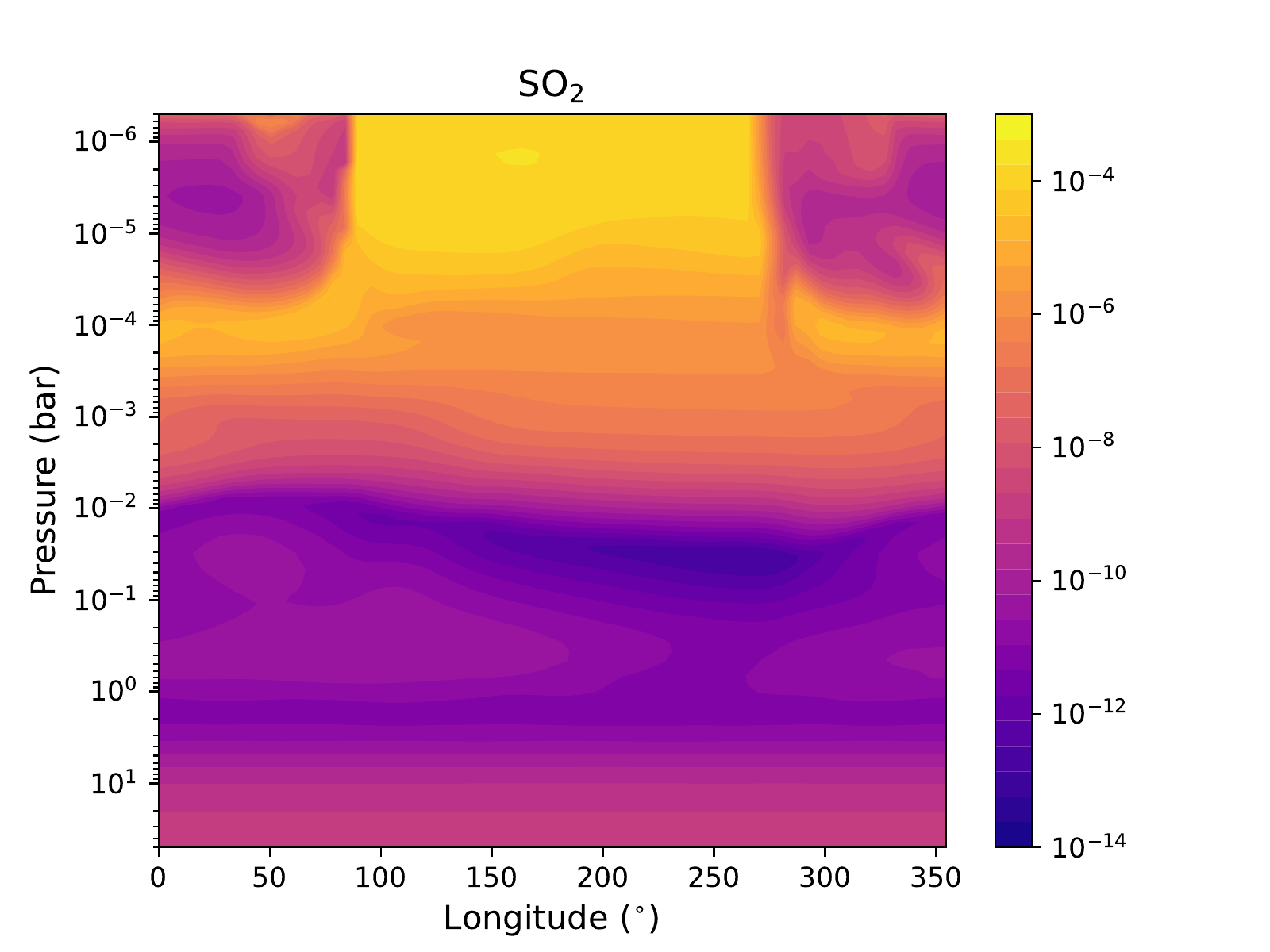}
\includegraphics[width=0.495\columnwidth]{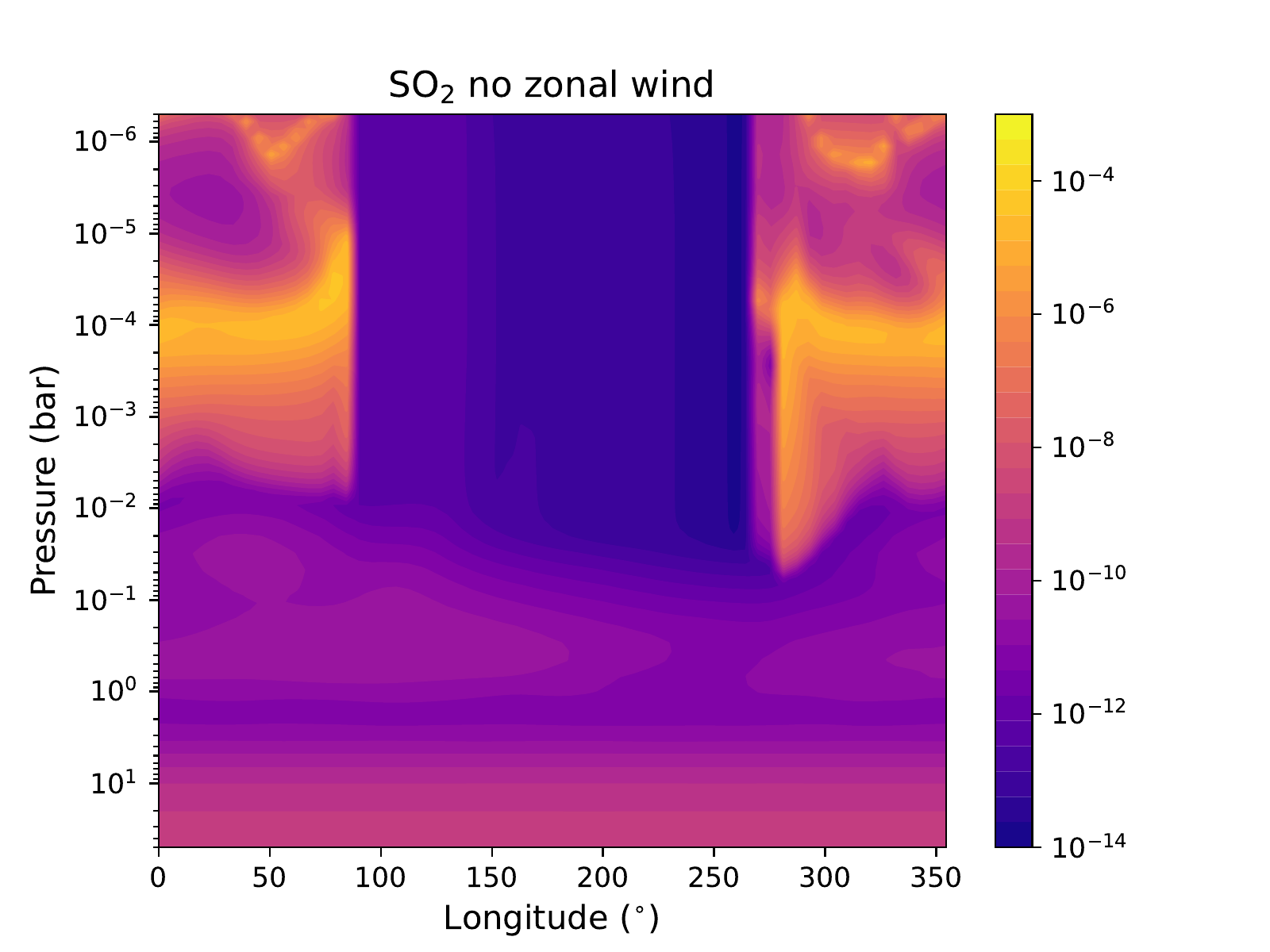}
\includegraphics[width=0.495\columnwidth]{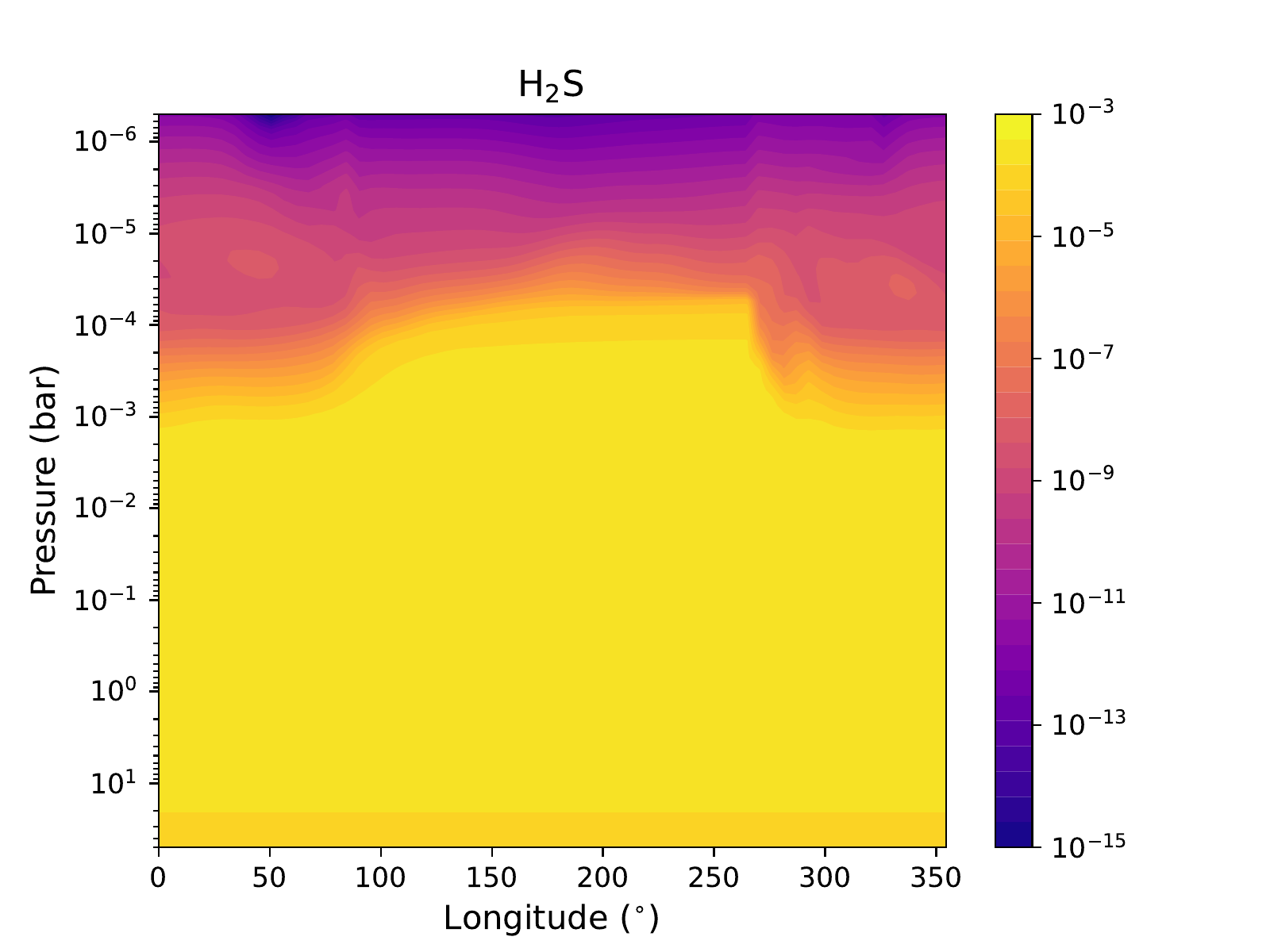}
\includegraphics[width=0.495\columnwidth]{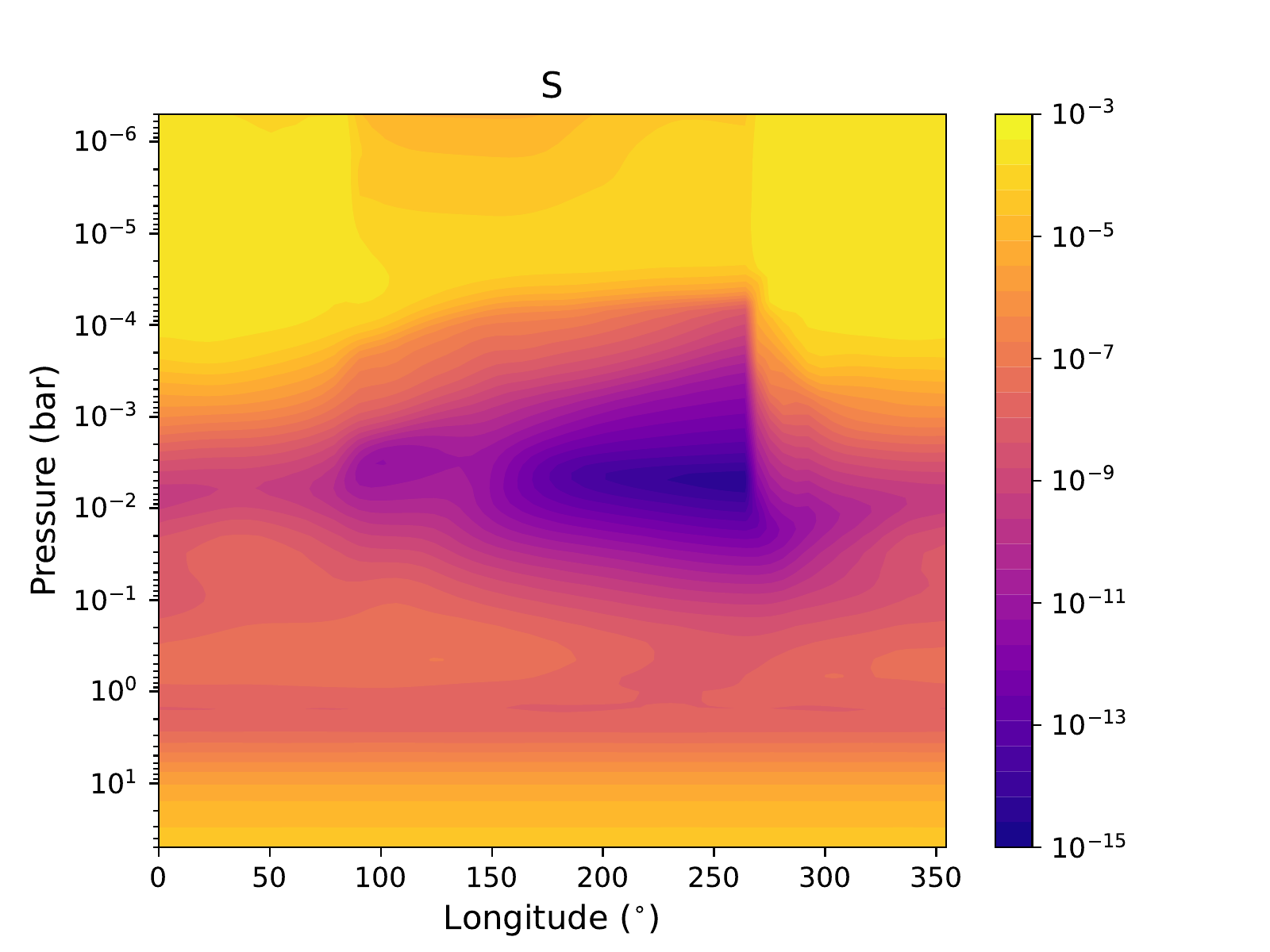}
\includegraphics[width=0.495\columnwidth]{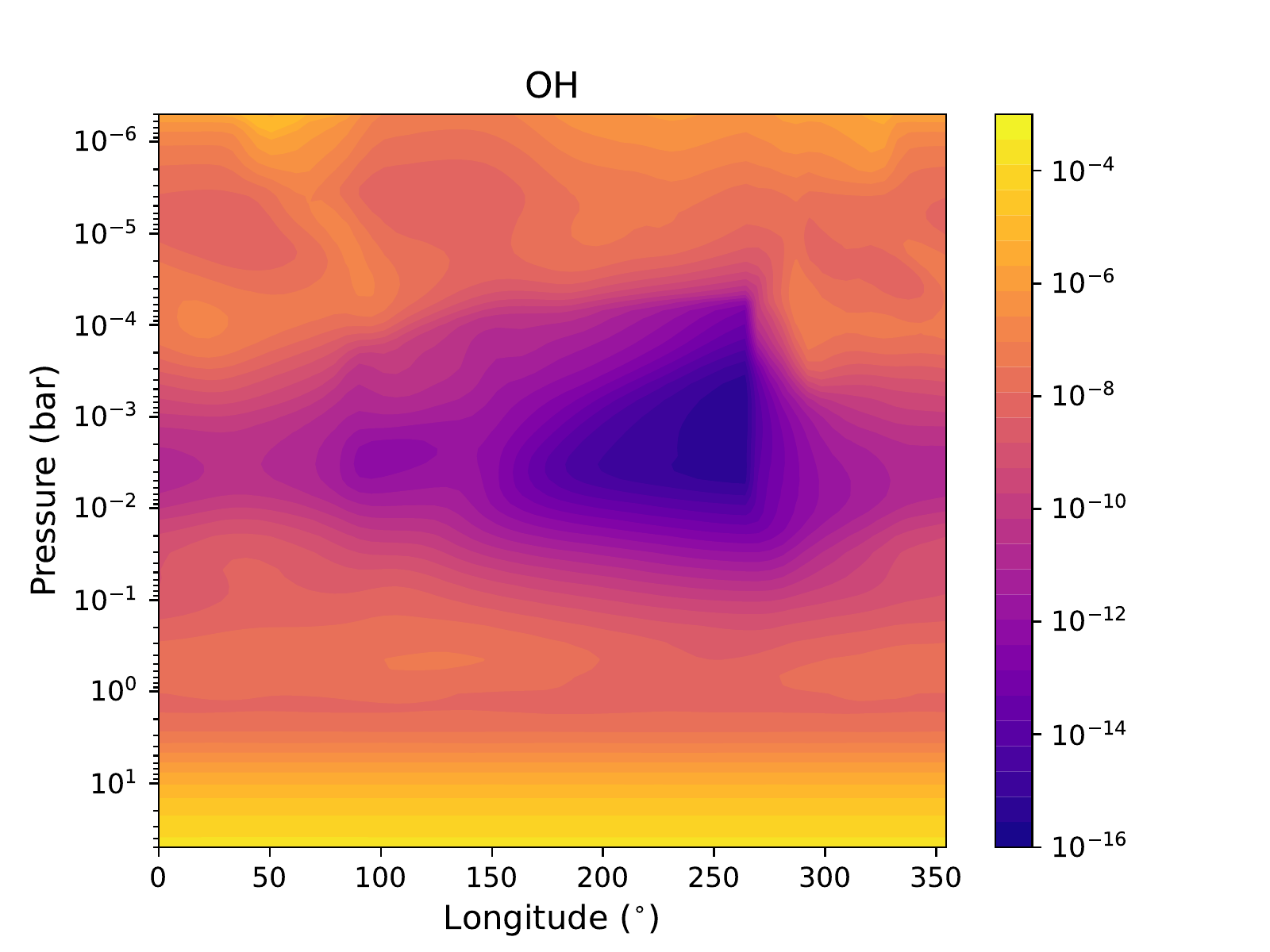}
\includegraphics[width=0.495\columnwidth]{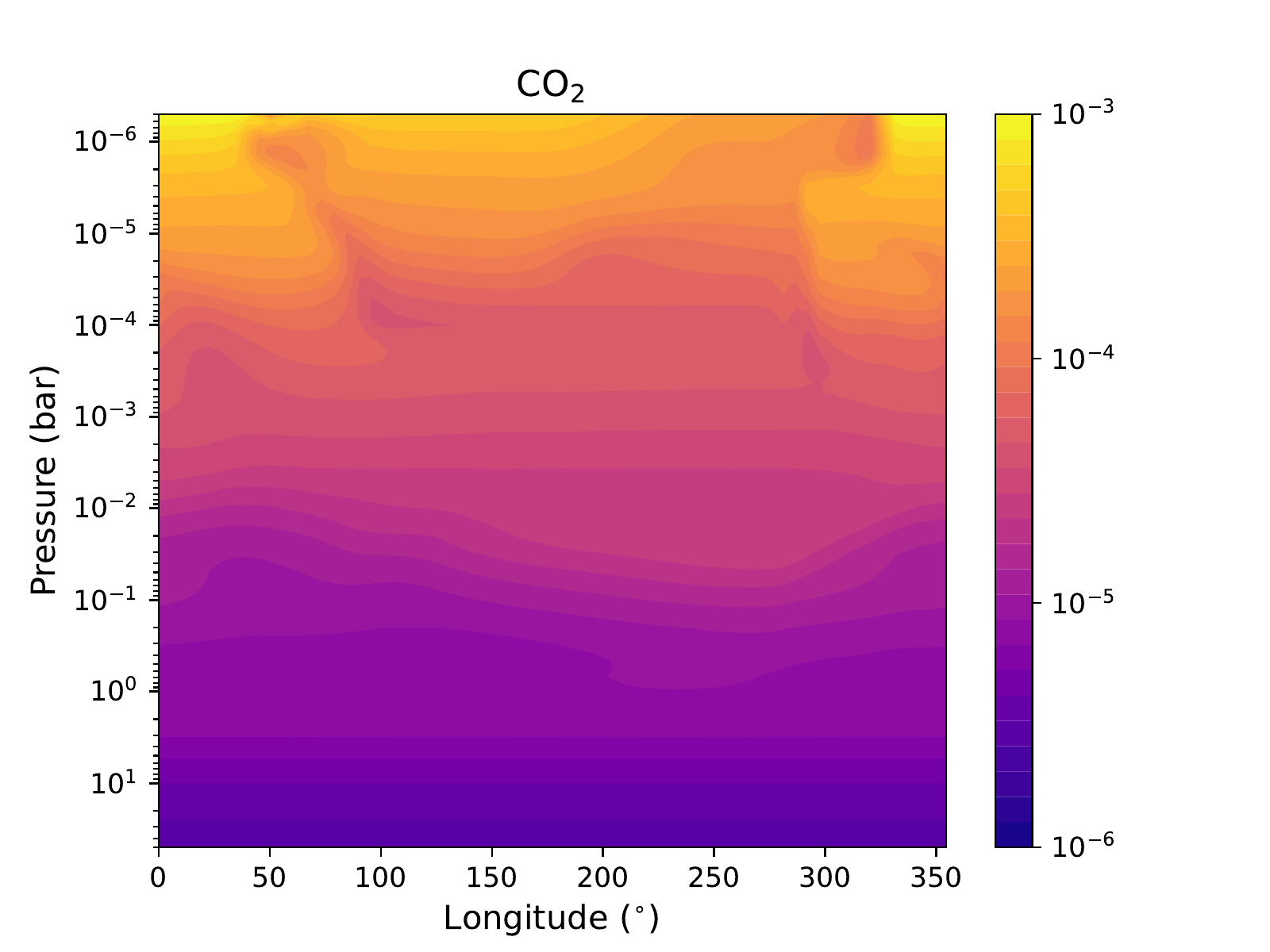}

\caption{The equatorial mixing-ratio distribution of several species participating in the sulfur cycle, along with \ce{CO2}, as a function of longitude and pressure on WASP-39b. The \ce{SO2} distribution when horizontal transport is omitted is shown in the upper right panel for comparison. The substellar point is located at 0$^{\circ}$ longitude.} 
\label{fig:2D_SO2}
\end{figure}

\begin{figure}[ht!]
\plottwo{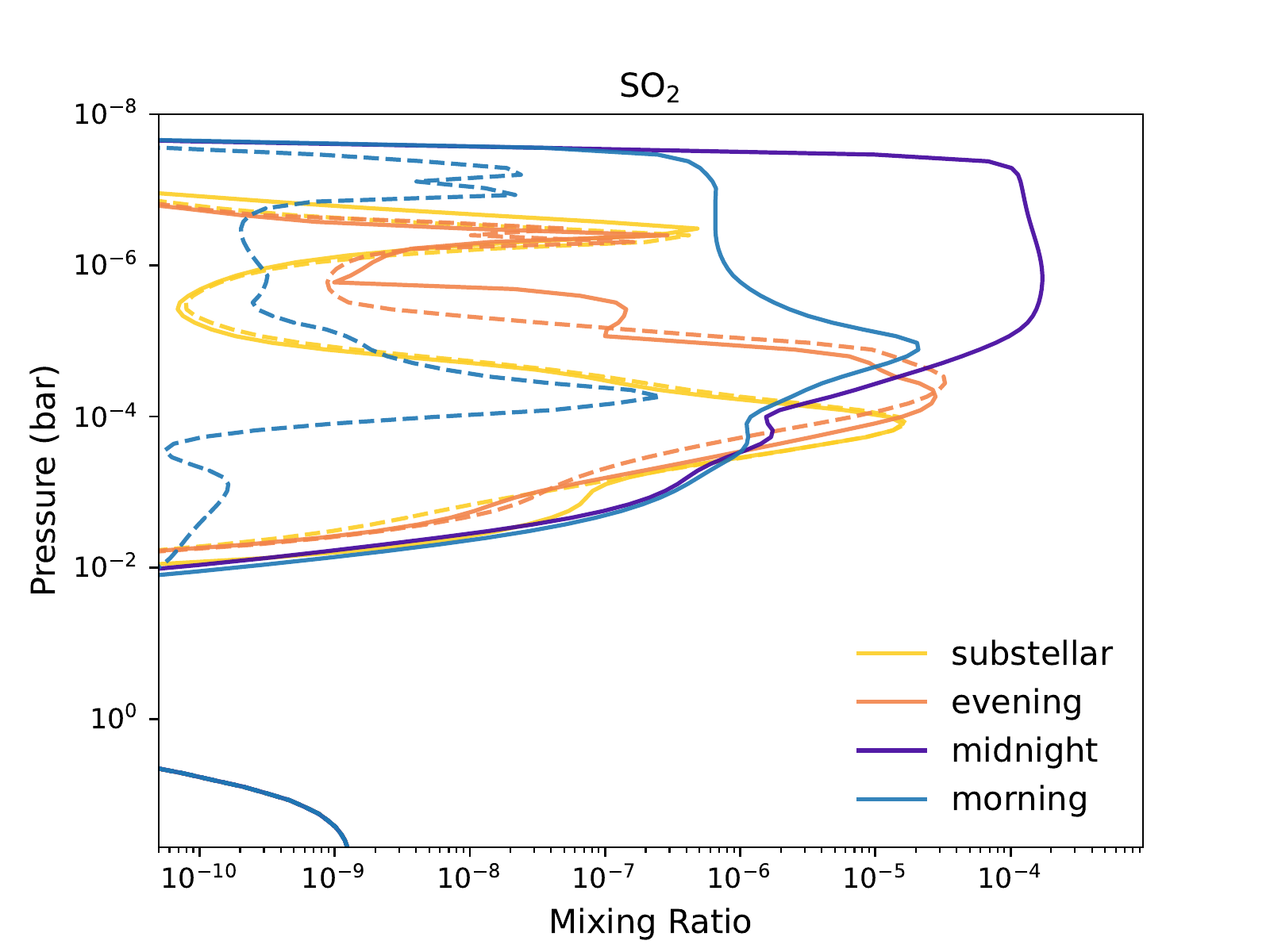}{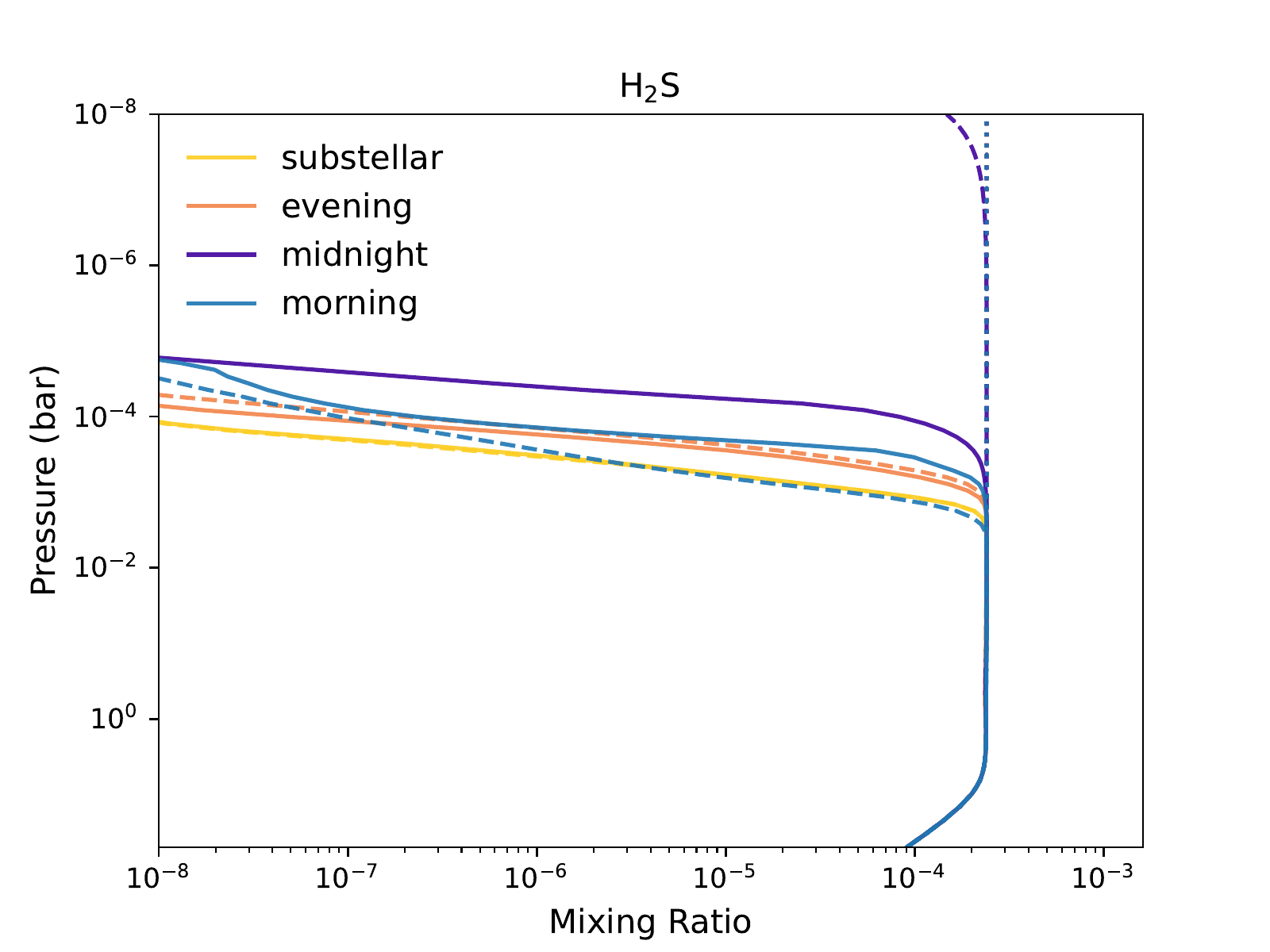}
\plottwo{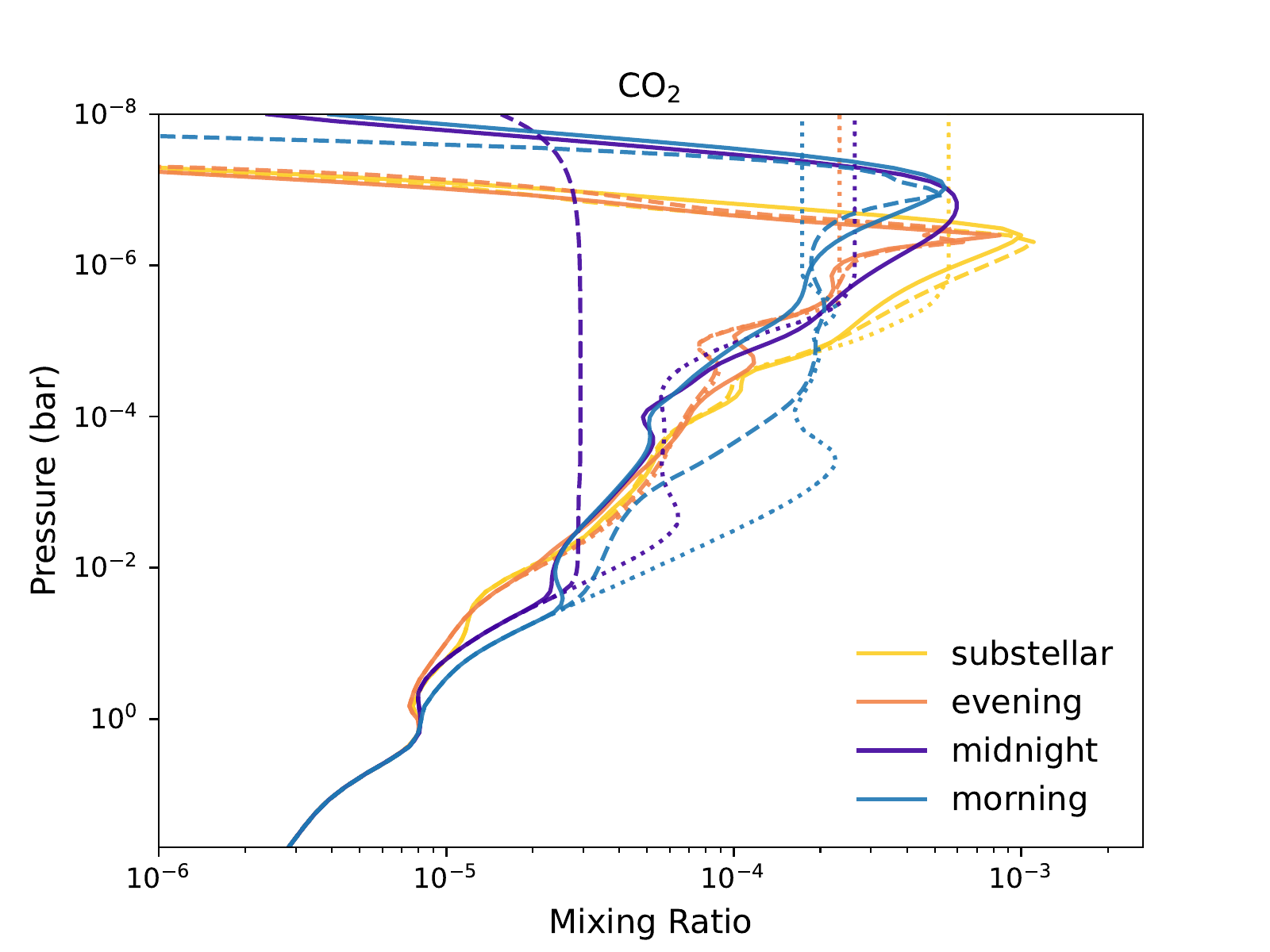}{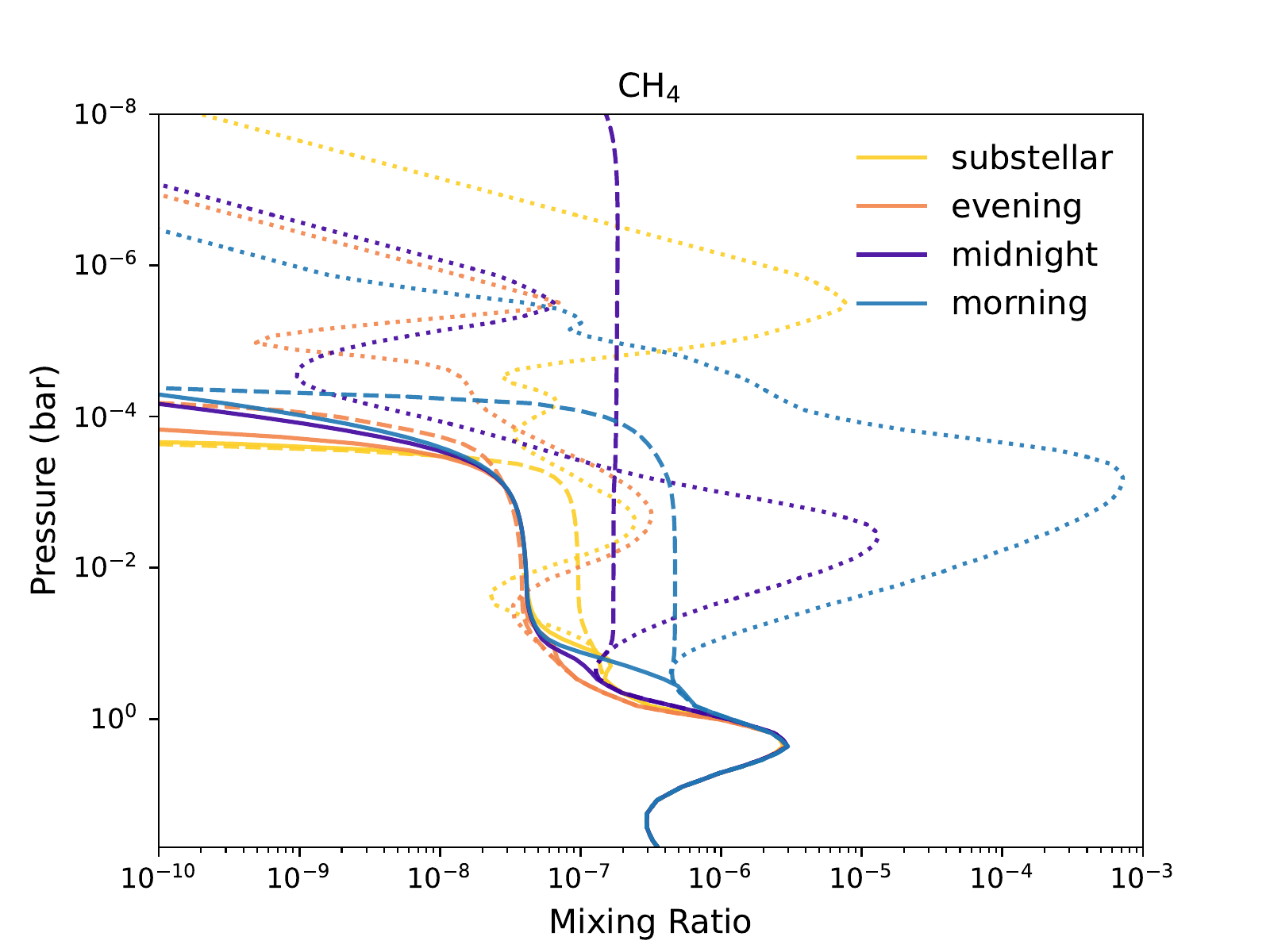}
\caption{The VMR profiles of \ce{SO2}, \ce{H2S}, \ce{CO2}, and \ce{CH4} at four different longitudinal locations: substellar point (0$^{\circ}$), evening (79$^{\circ}$), midnight (180$^{\circ}$), and morning (259$^{\circ}$). Results are compared to the vertical-mixing case where horizontal transport is excluded (dashed) and chemical equilibrium (dotted) to provide insights into the role of horizontal transport in shaping limb differences.}
\label{fig:1D}
\end{figure} 

\begin{figure}[ht!]
\includegraphics[width=0.5\columnwidth]{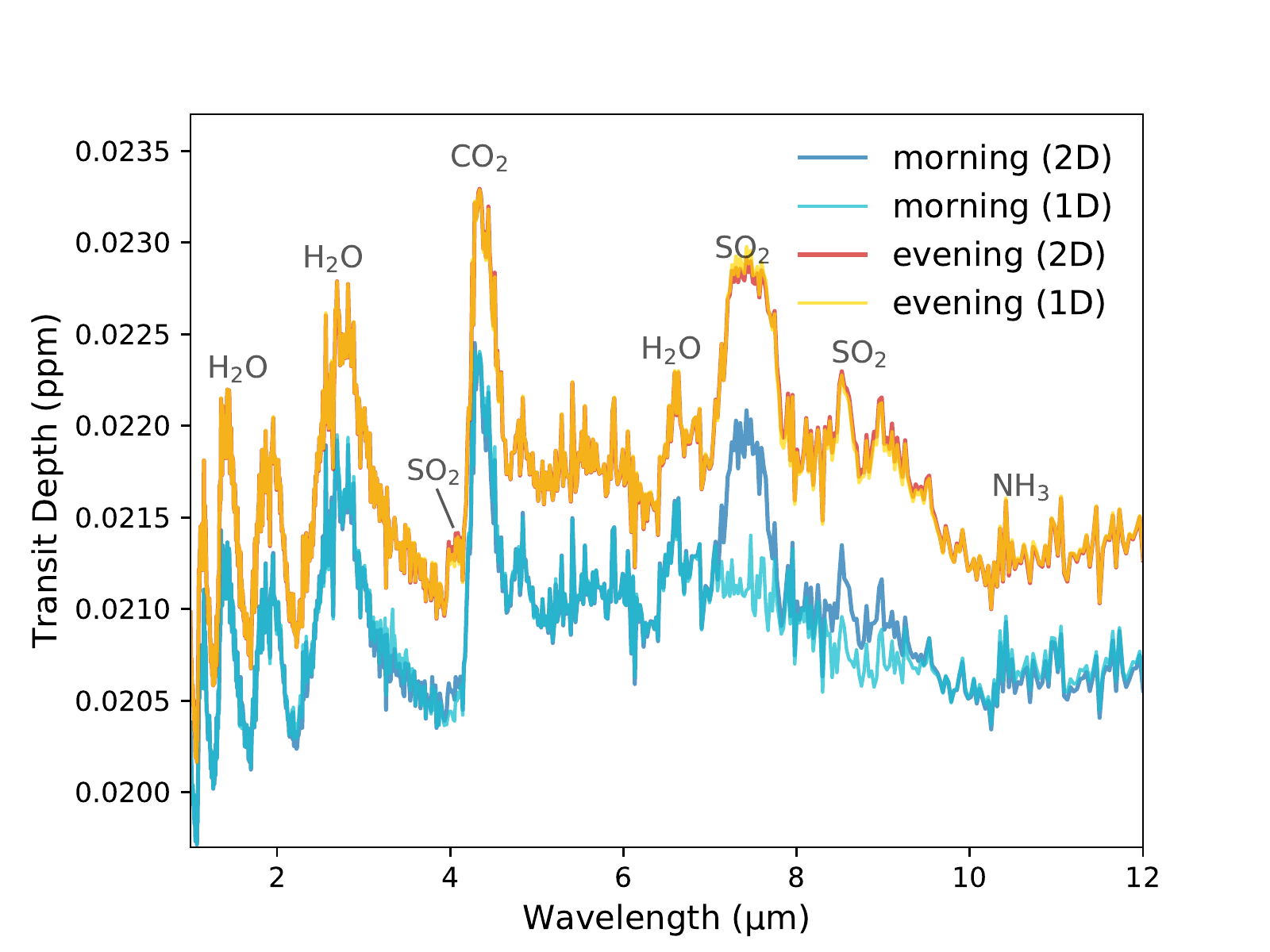}
\includegraphics[width=0.5\columnwidth]{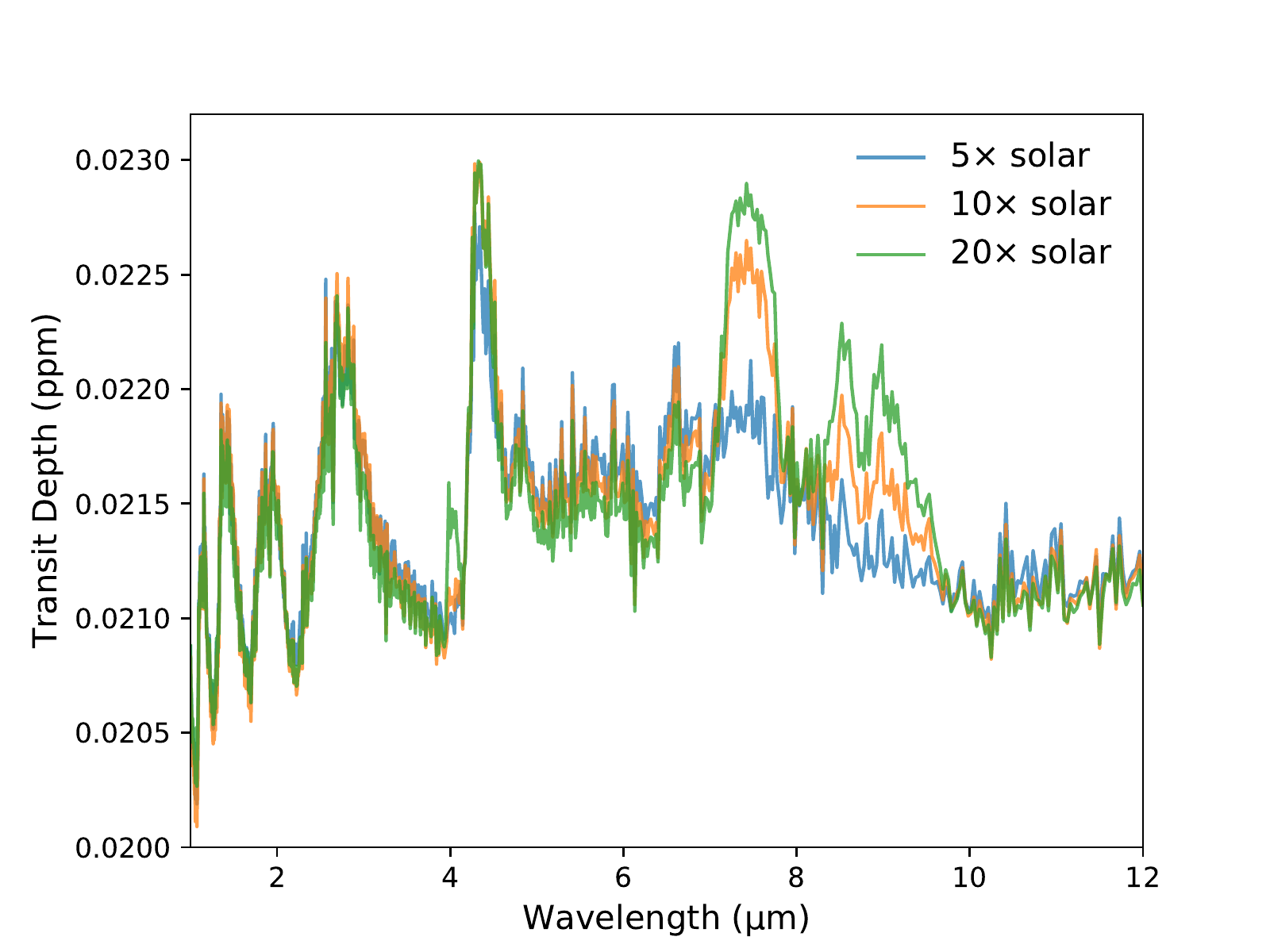}
\caption{The left panel shows the synthetic transmission spectra of the evening and morning limbs of WASP-39b with a resolution of R = 300. The atmospheric structure and chemical composition are based on 2D VULCAN, incorporating temperature and wind profiles obtained from the Exo-FMS GCM. The models excluding horizontal transport are labelled 1D for comparison. The right panel shows the morning-evening averaged transmission spectra when assuming different atmospheric metallicities within 2D VULCAN.}
\label{fig:transmission}  
\end{figure}

\section{Methods}
The equatorial jet induced by day-night thermal forcing on synchronously rotating planets has been extensively studied by numerical simulations and confirmed with phase curve observations (see \cite{Showman2020} for a review). This fast wind plays a crucial role in transporting chemical abundances and cloud particles horizontally across the planet. In this study, we utilize the benefits of 2D models that preserve detailed chemical kinetics to investigate the effects of day-night transport. Our 2D models adopt the atmospheric structure at the equatorial region from the same 3D WASP-39b atmospheric simulation used in \cite{Tsai2023b}. 

\subsection{3D GCM}\label{sec:GCM}
The Exo-FMS GCM \citep{Lee2021} employs a correlated-k radiative transfer scheme, assuming cloud-free and thermochemical equilibrium under 10$\times$ solar metallicity for WASP-39b. We used a resolution of C48 (approximately 192 $\times$ 96 in longitude $\times$ latitude) for the cube-sphere grid. The adopted parameters for our WASP-39b GCM are listed in Appendix Table \ref{tab:GCM_parameters} with further details described in Methods section in \cite{Tsai2023b}. To supply input for the 2D chemistry and cloud models, the temperature and wind fields are averaged over the equatorial region of $\pm$ 30$^{\circ}$, where the circulation is dominated by a zonal jet and can be captured with a 2D framework. The equatorial temperature and winds can be found in the right panel of Appendix Figure \ref{fig:tau_dyn}. Since the 2D photochemical model requires a higher upper boundary than the GCM, the temperatures and winds at the top GCM boundary of about 10$^{-6}$ bar are extrapolated as constant values up to 10$^{-8}$ bar for the 2D model.

The equatorial region is divided into 64 longitude columns. We have performed additional tests with 16 and 32 longitude columns and found minimal differences across various zonal resolutions. We incorporate the longitudinal dependence of vertical mixing, instead of using the global fitting expression applied in \cite{Tsai2023b}. We use the root-mean-square of the vertical wind velocity ($w_{\textrm{rms}}$) to estimate the eddy diffusion coefficients ($K_{\textrm{zz}}$) from mixing length theory: $K_{\textrm{zz}}$ = 0.1$\times w_{\textrm{rms}} H$, where 0.1$\times H$ (scale height) is taken to be the characteristic length \citep{Smith1998,Parmentier2013}. The first panel of Appendix Figure \ref{fig:tau_dyn} illustrates the timescales due to horizontal transport and vertical mixing. Owing to the inflated scale height of WASP-39b, the vertical mixing timescale appears to be slightly longer than typical hot Jupiters \citep{Parmentier2013,Xi2018,Komacek2019}. Our GCM modeling suggests that transport via horizontal winds is expected to be the dominant mixing process in the atmosphere of WASP-39b.  



\subsection{2D photochemistry}\label{sec:2D-chem}
We apply the 2D photochemical model VULCAN (2D VULCAN) to account for both vertical mixing and horizontal transport of gas-phase constituents on an equatorial plane. We note that the actual 2D pressure-longitude grid provides us with the flexibility to adopt any zonal wind profiles derived from the GCM, whereas the pseudo-2D (rotating 1D) model \citep[e.g.,][]{Agundez2014,Moses2021,Venot2020,Baeyens2022} is limited to a uniform zonal flow. 

The 2D model solves the continuity equations including horizontal transport
\begin{linenomath*}
\begin{equation}
\frac{\partial n_i}{\partial t} = {\cal P}_i - {\cal L}_i - \frac{\partial \phi_{i,z}}{\partial z} - \frac{\partial \phi_{i,x}}{\partial x},
\label{eq:master}
\end{equation}
\end{linenomath*}
where $n_i$ is the number density (cm$^{-3}$) of species $i$, ${\cal P}_i$ and ${\cal L}_i$ are the chemical production and loss rates (cm$^{-3}$ s$^{-1}$) of species $i$, and $\phi_{i,z}$, $\phi_{i,x}$ are the vertical and horizontal transport flux (cm$^{-2}$ s$^{-1}$), respectively. The vertical flux includes molecular diffusion and eddy diffusion, expressed as 
\begin{equation}
\phi_{i,z} = - K_{\textrm{zz}} N \frac{\partial X_{i,z}}{\partial z} - D_{i,z} \left[ \frac{\partial n_{i,z}}{\partial z} + n_{i,z}(\frac{1}{H_{i,z}} + \frac{1+\alpha_T}{T}\frac{d T}{d z}) \right],
\label{eq:phiz}
\end{equation}
where $D_{i,z}$ is the molecular diffusion coefficient, $N$ is the total number density, and $\alpha_T$ is the thermal diffusion factor. The horizontal flux comprises advection by the pressure-dependent zonal winds ($v_x$), expressed as
\begin{equation}
\phi_{i,x} = - n_{i,x} v_x.
\label{eq:phix}
\end{equation}
In this study, we did not consider vertical advection transport \citep{Tsai2021}. For horizontal advection transport, we use a first-order upwind difference scheme evaluated at isobaric surfaces, where the local concentration is affected by the upstream grid only (same as the vertical advection transport described in \cite{Tsai2021}). The vertical eddy diffusion coefficients and zonal winds are derived from the GCM, as described in Section \ref{sec:GCM}. The GCM-derived temperature and wind fields remain fixed without incorporating the feedback from compositional changes. When varying atmospheric metallicity in the 2D VULCAN model later, we adopted identical GCM outputs based on 10 $\times$ solar metallicity. The construction and validation of 2D VULCAN are presented in detail in a separate paper(submitted).

\section{Results: Circulation-modulated gas-phase photochemistry}\label{sec:2d_vul_results}

JWST observations and modeling efforts have confirmed that \ce{SO2} can be produced from \ce{H2S} in the \ce{H2}-based atmosphere of WASP-39b, through the aid of atomic hydrogen (H) and hydroxyl radicals (OH) generated by \ce{H2O} photolysis, followed by other kinetic reactions. Figure \ref{fig:2D_SO2} illustrates the distributions of several species involved in the \ce{SO2} production in the equatorial region of WASP-39b computed by our 2D model.

The first remarkable feature is the accumulation of \ce{SO2} in the nightside upper atmosphere, with a peak volume mixing ratio (VMR) of 100 ppm above 10$^{-5}$ bar. The high abundance of \ce{SO2} in the nightside upper atmosphere might seem unintuitive at first, since \ce{SO2} is a photochemical product originating from the dayside under the influence of stellar UV. Without zonal wind (top right panel of Figure \ref{fig:2D_SO2}), nightside \ce{SO2} would have remained negligible. In fact, we find that sulfur can still be liberated from \ce{H2S} on the nightside through advecting the dayside-originated H and OH. The same oxidation pathways as elucidated in \cite{Tsai2023b} proceed on the nightside. The only difference is the photodissociation of \ce{SO2} responsible for the destruction of \ce{SO2} in the upper atmosphere is absent on the nightside, which explains the buildup of \ce{SO2} in the nightside upper atmosphere. Note that at these high altitudes, the photochemical destruction of \ce{SO2} is faster than the zonal-wind transport such that the high-altitude \ce{SO2} produced on the nightside does not spread globally.
 

To further elucidate the role of zonal transport, we performed additional tests to determine the lifetime of \ce{SO2} in the absence of replenishment from the dayside. Appendix Figure \ref{fig:SO2_evo} shows that nightside \ce{SO2} would dissipate into the deep atmosphere and reform \ce{H2S} over the timescale of $\sim$ 10--100 days. To sustain \ce{SO2} on the nightside, horizontal transport must act at a faster rate than this lifetime of \ce{SO2}. According to Appendix Figure \ref{fig:tau_dyn}, the timescale of horizontal transport above 1 bar is approximately 1 day, which confirms that the zonal wind can continuously resupply the radicals needed to produce \ce{SO2} on the nightside. 

To directly evaluate the compositional inhomogeneity across the planet, Figure \ref{fig:1D} presents the vertical distribution of \ce{SO2}, \ce{H2S}, \ce{CO2}, and \ce{CH4} at four representative longitudinal locations. The differences between the morning and evening limbs can also be examined in comparison with the cases of vertical mixing (excluding horizontal transport) and chemical equilibrium. The \ce{SO2} distribution on the dayside hemisphere is not significantly affected by the zonal wind. Instead, photochemical production and destruction dominate.  As previously discussed, \ce{SO2} around the antistellar point builds up in the upper atmosphere above 0.1 mbar via zonal transport, exceeding the dayside abundances. Without horizontal transport, the abundance of \ce{SO2} is orders of magnitude lower on the morning limb. Note that in contrast, \ce{SO2} is predicted to be more abundant on the morning limb in \cite{Tsai2023b}. This difference is because in our 2D model grid, the local temperature around the morning terminator is cooler than the morning-limb-averaged profile (over 20$^{\circ}$ longitudinal region) in \citep{Tsai2023b}, which suppresses \ce{SO2} production. The introduction of transport from the nightside results in a reduction of \ce{SO2} below 1 mbar on the morning limb, bringing the morning and evening \ce{SO2} abundances close to each other.


On the other hand, \ce{CH4} is controlled by vertical quenching from the deeper regions rather than being of photochemical origin like \ce{SO2} in the upper atmosphere. In the absence of horizontal transport, the vertically-quenched \ce{CH4} between 0.1 and 10$^{-4}$ bar exhibits a notable compositional gradient between longitudes, where the photospheric \ce{CH4} VMR (i.e., at 1 -- 10$^{-4}$ bar) on the cooler morning limb is about an order of magnitude higher than that on the evening limb. This difference results from the different atmospheric temperatures at the vertical quench point, with the morning limb being cooler and thus supporting more \ce{CH4}.  However, when horizontal transport is introduced, \ce{CH4} is readily homogenized and ends up with a VMR that more closely matches the warmer evening profile everywhere across the planet. Lastly, \ce{CO2} exhibits a fairly uniform distribution close to the equilibrium abundance across the planet. An intriguing feature is that 1D models that neglect horizontal transport appear to slightly underestimate the \ce{CO2} abundance on the nightside due to vertical quenching, which would be replenished when horizontal transport is included.       






\section{Features in the synthetic spectra}
Here, we will first examine the observational implications of the computed gas-phase constituents on WASP-39b. Transmission spectral components at the transit ingress and egress can be separated to individually probe the morning and evening limbs \citep{Espinoza2021,Grant2023}. Thermal emission observations at different orbital phases further reveal the spatial variation across the planet \citep{Stevenson2014,Mendonca2018,Chubb2022,Kempton2023}. While most previous studies focused on temperature variations, our integrated approach enables us to examine the effects of chemical variations regulated by both vertical and horizontal transport. 

The atmospheric structure and composition distributions of WASP-39b computed by our 3D GCM and 2D atmosphere models are put together to generate planetary spectra. We use PLATON \citep{Zhang2019,Zhang2020} to make transmission spectra and HELIOS \citep{Malik2019} to make emission spectra. In this work, we will simply assume cloud-free when making the spectra and include clouds in the follow-up paper.



\subsection{Transmission spectra}
The left panel of Figure \ref{fig:transmission} shows the transmission spectra of morning and evening limbs separately. The spectra on both limbs both show \ce{CO2} features between 4.2 and 4.5 $\mu m$, \ce{H2O} features between 1 and 3 $\mu m$, and \ce{SO2} features at 4.1 and 7--9 $\mu m$. While horizontal transport has negligible effects on the evening limb, it increases \ce{SO2} to detectable levels in the morning limb, compared to 1D prediction which only considers vertical mixing (discussed in Section \ref{sec:2d_vul_results}), as seen in the 7-$\mu m$ feature. Nonetheless, with additional 2D VULCAN runs with 5 and 20 times atmospheric metallicities, we find the sensitivity of \ce{SO2} to metallicity pointed out in \cite{Tsai2023b} remains unchanged, as depicted in the right panel of Figure \ref{fig:transmission}. The main difference between the two limbs lies in the smaller scale height on the cooler morning limb driven by temperature. Conversely, the compositional differences between the two limbs are largely homogenized by the zonal wind. 




\subsection{Emission spectra}
\begin{figure}[htp!]
\includegraphics[width=0.333\columnwidth]{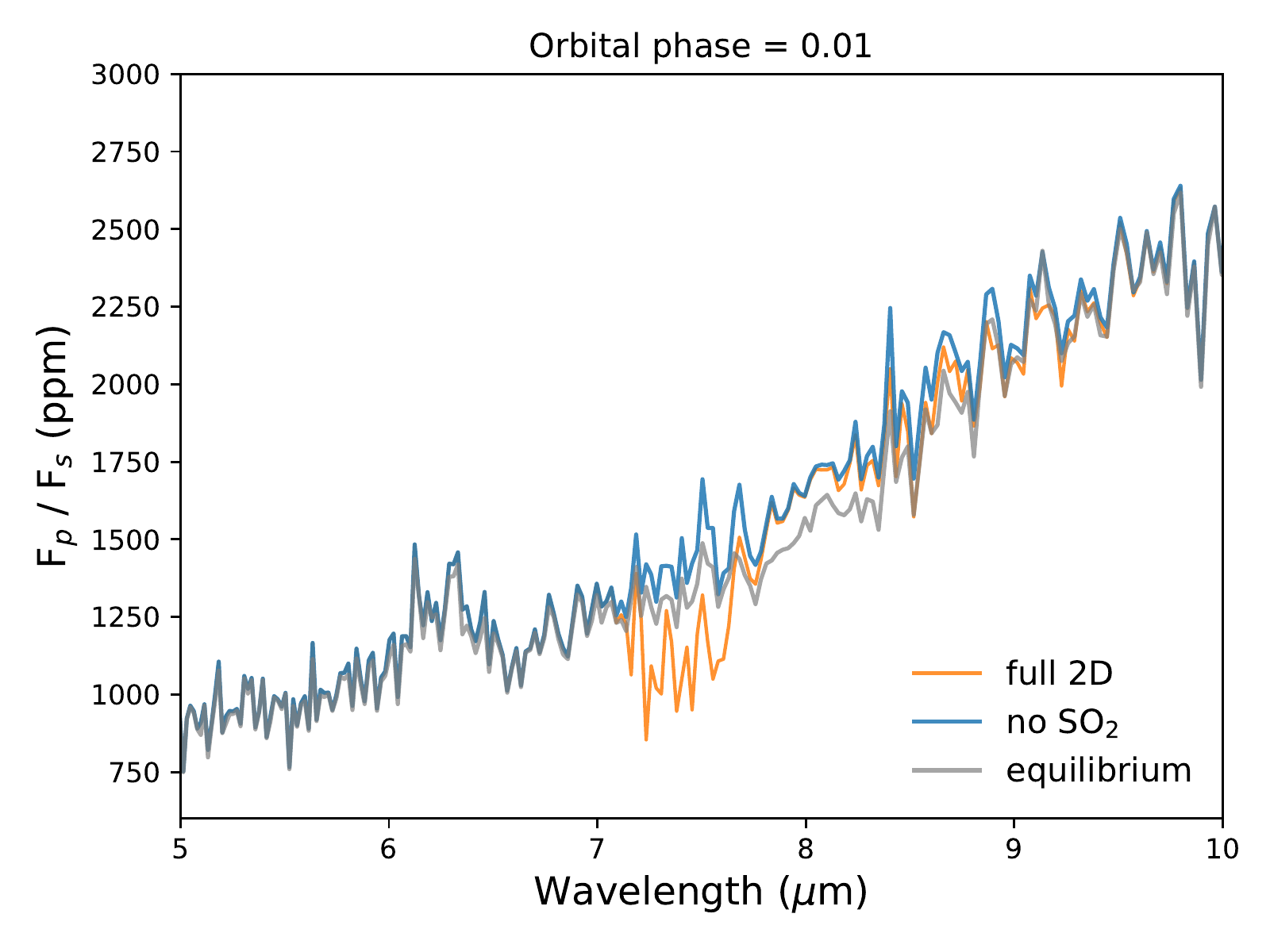}
\includegraphics[width=0.333\columnwidth]{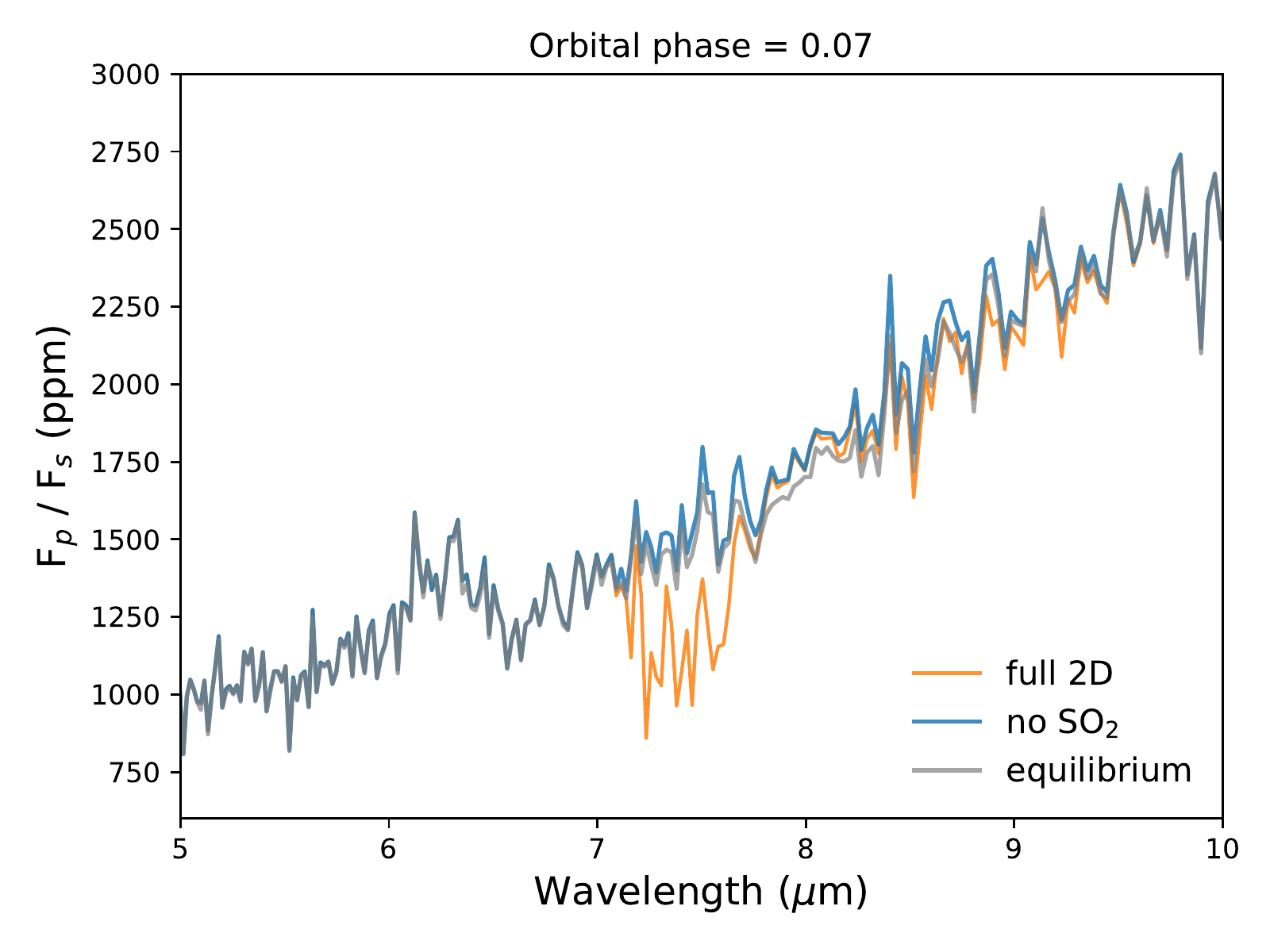}
\includegraphics[width=0.333\columnwidth]{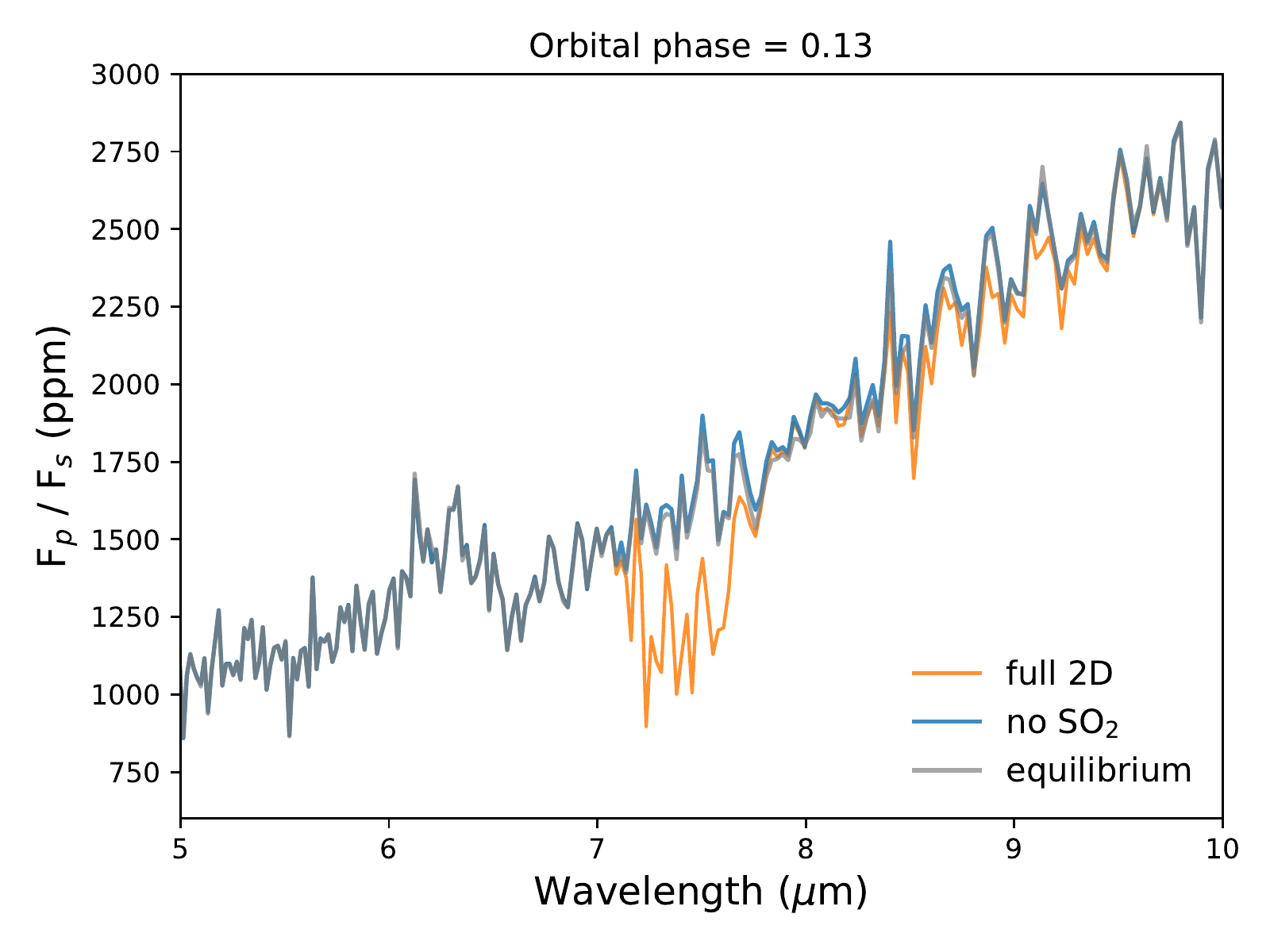}
\includegraphics[width=0.333\columnwidth]{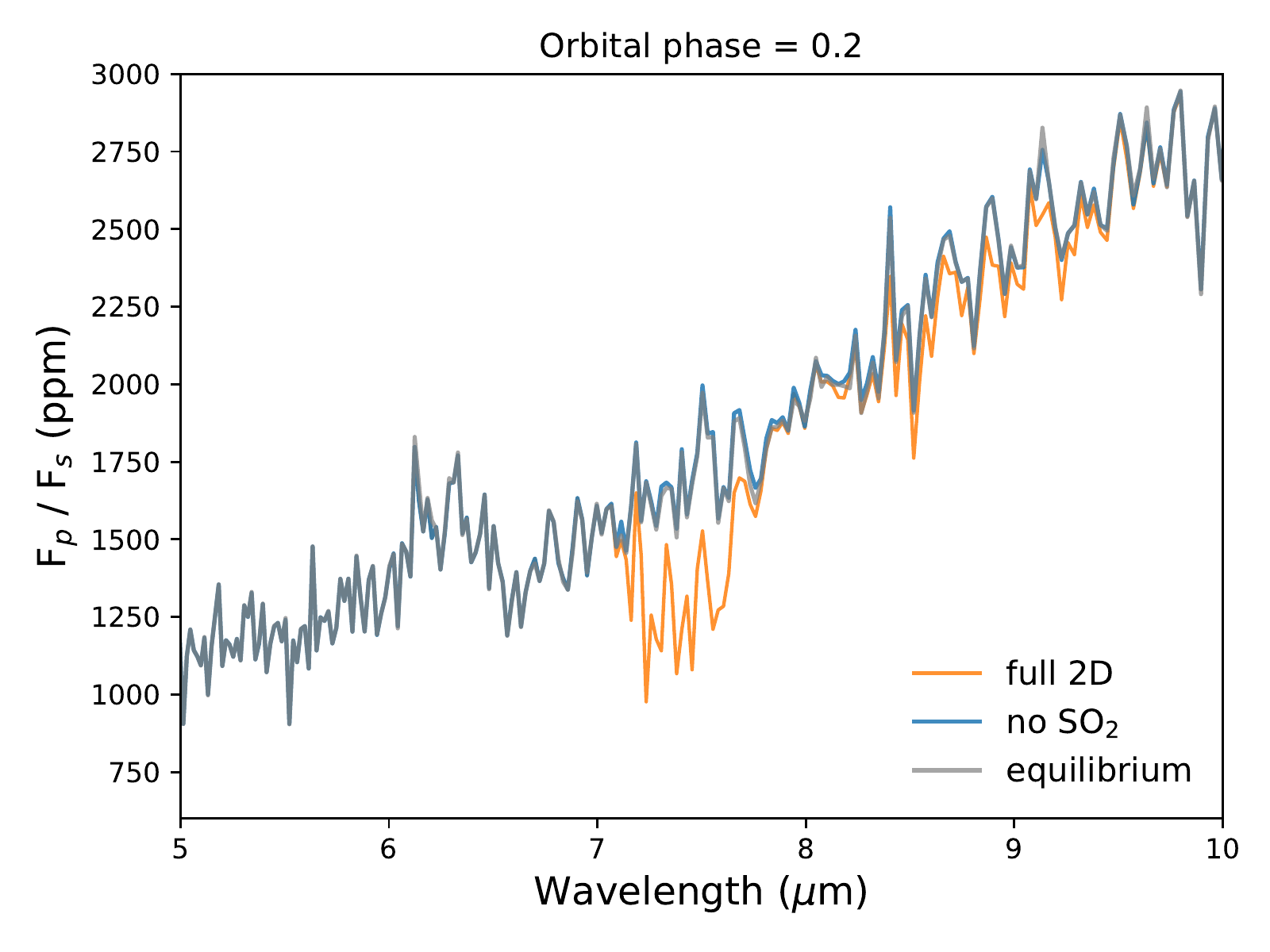}
\includegraphics[width=0.333\columnwidth]{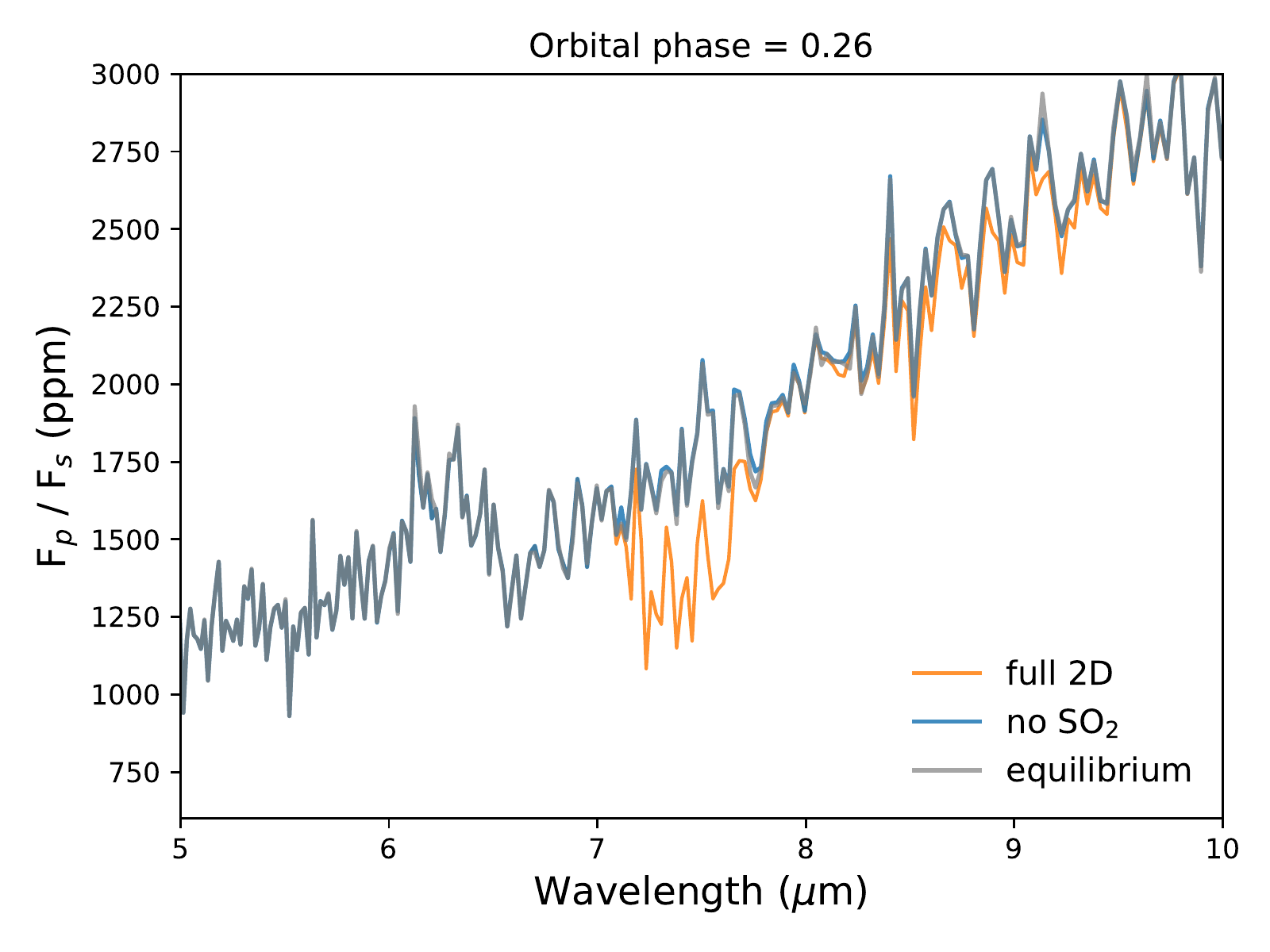}
\includegraphics[width=0.333\columnwidth]{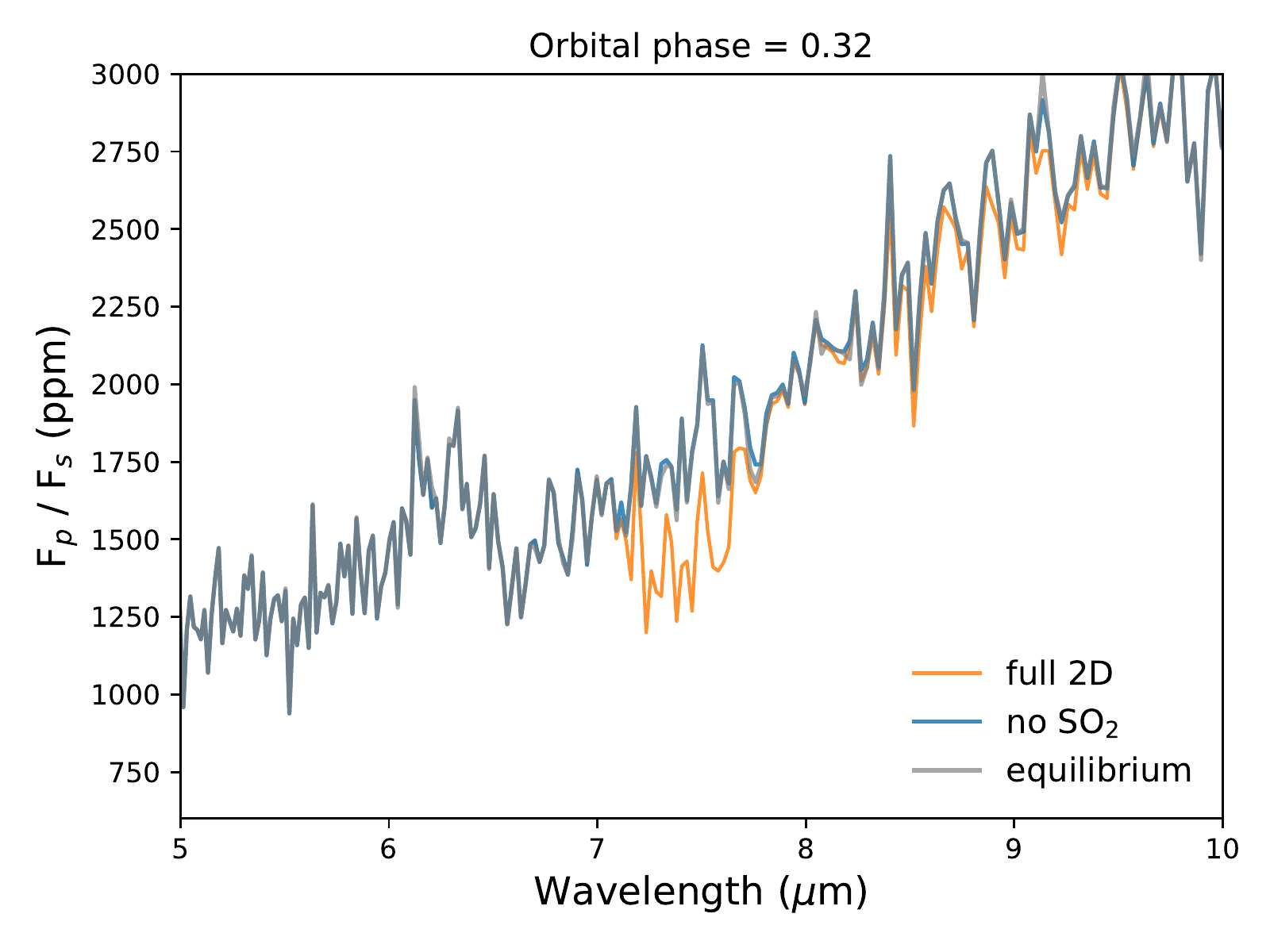}
\includegraphics[width=0.333\columnwidth]{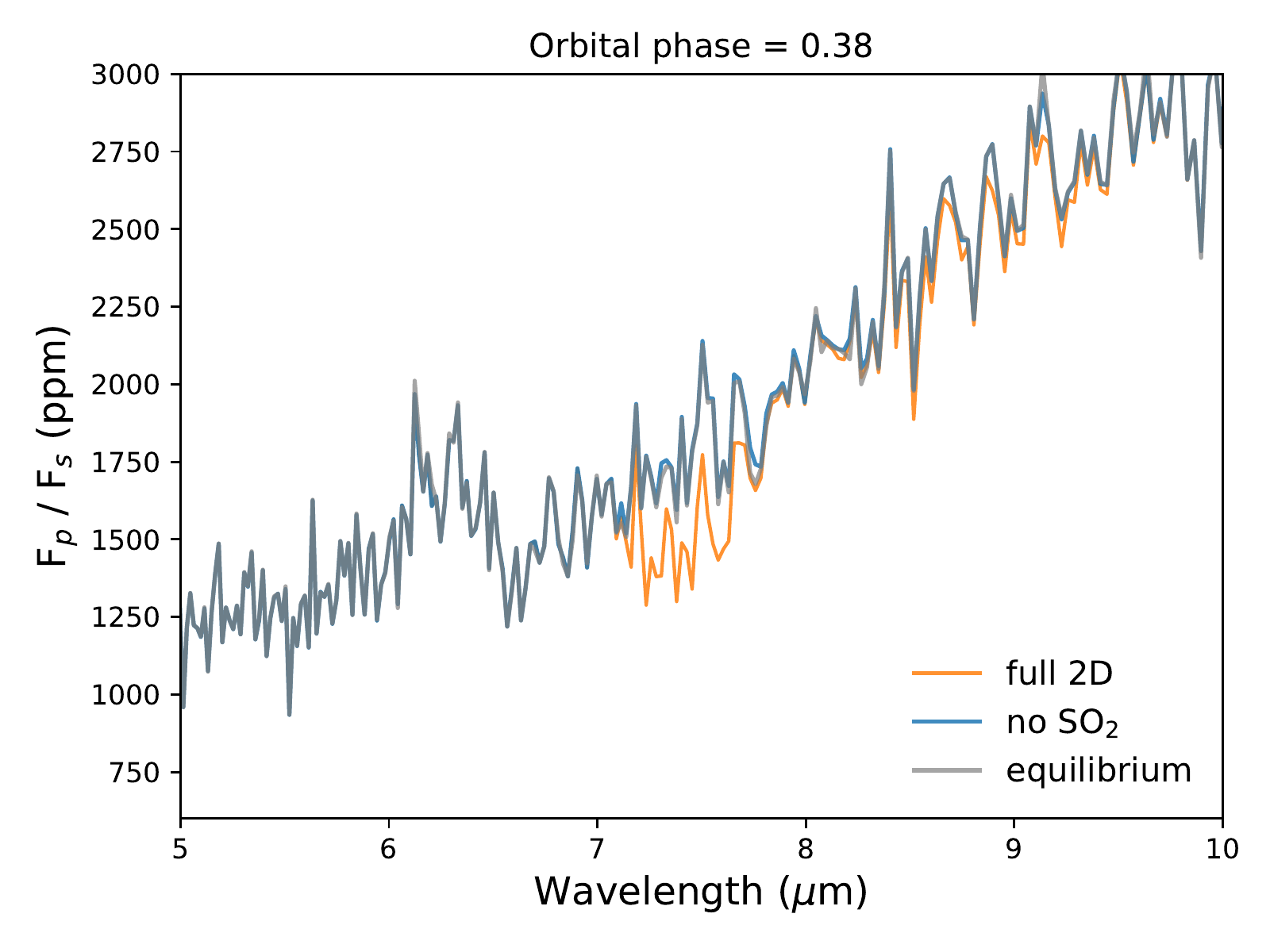}
\includegraphics[width=0.333\columnwidth]{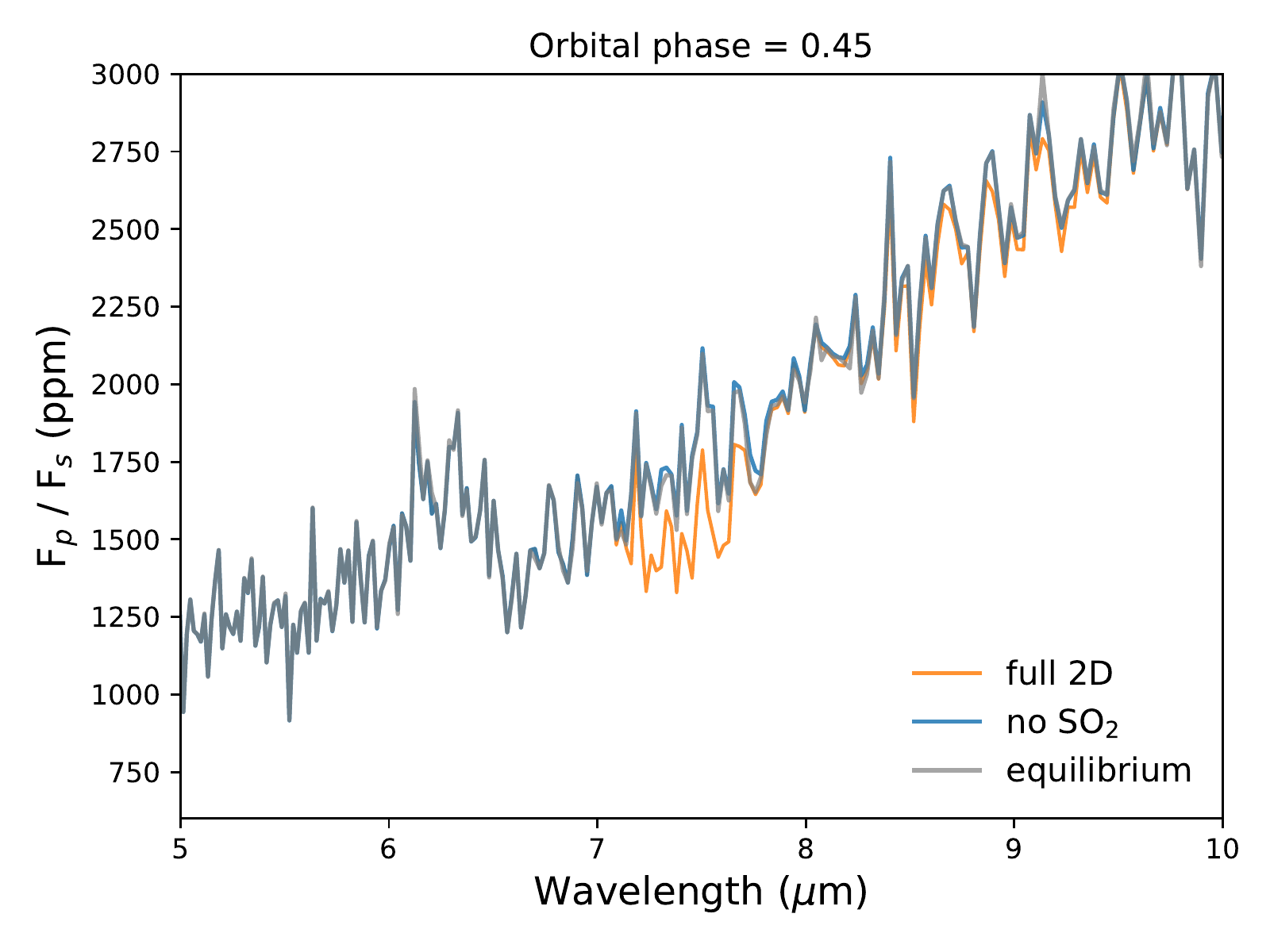}
\includegraphics[width=0.333\columnwidth]{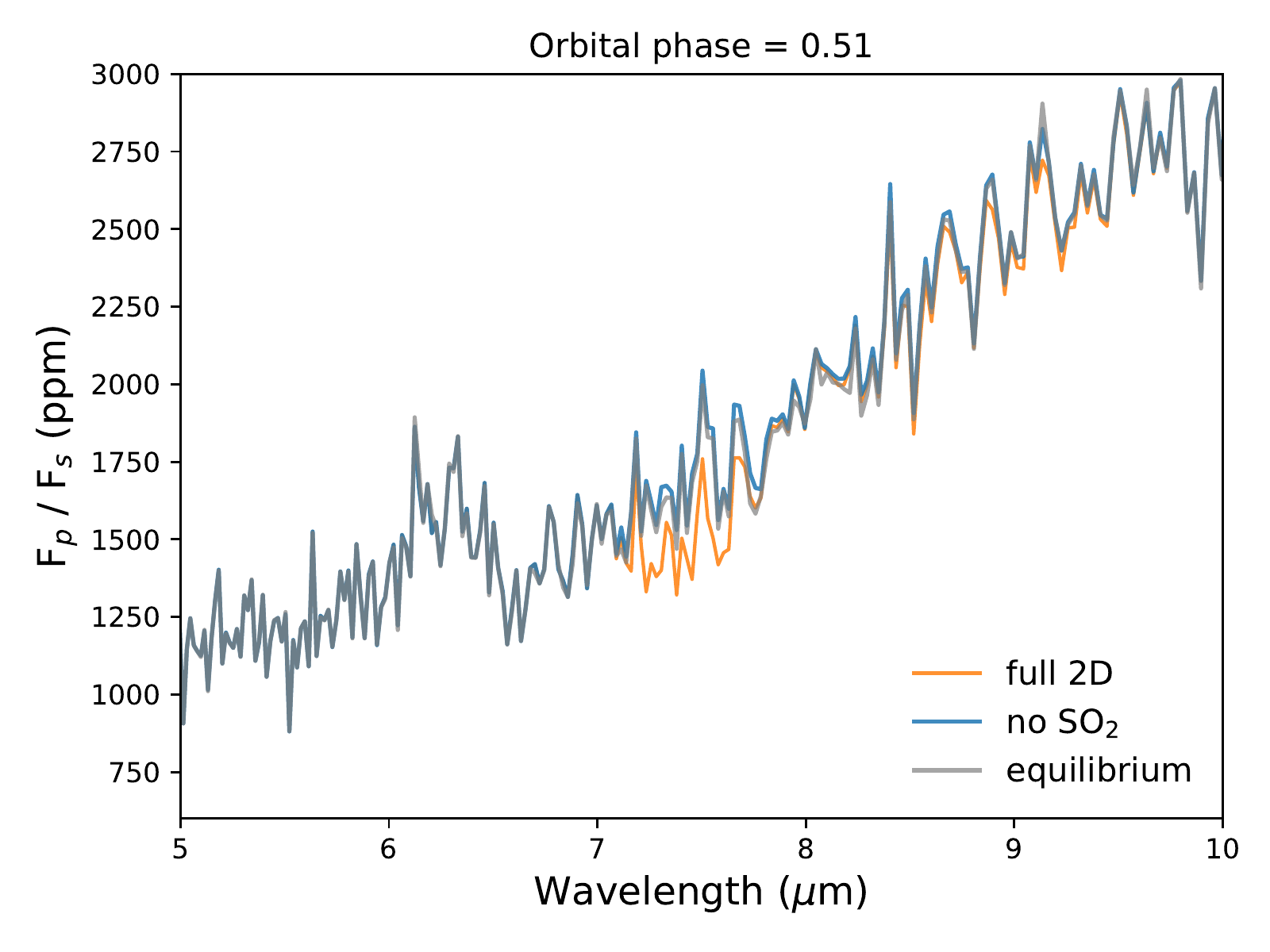}
\includegraphics[width=0.333\columnwidth]{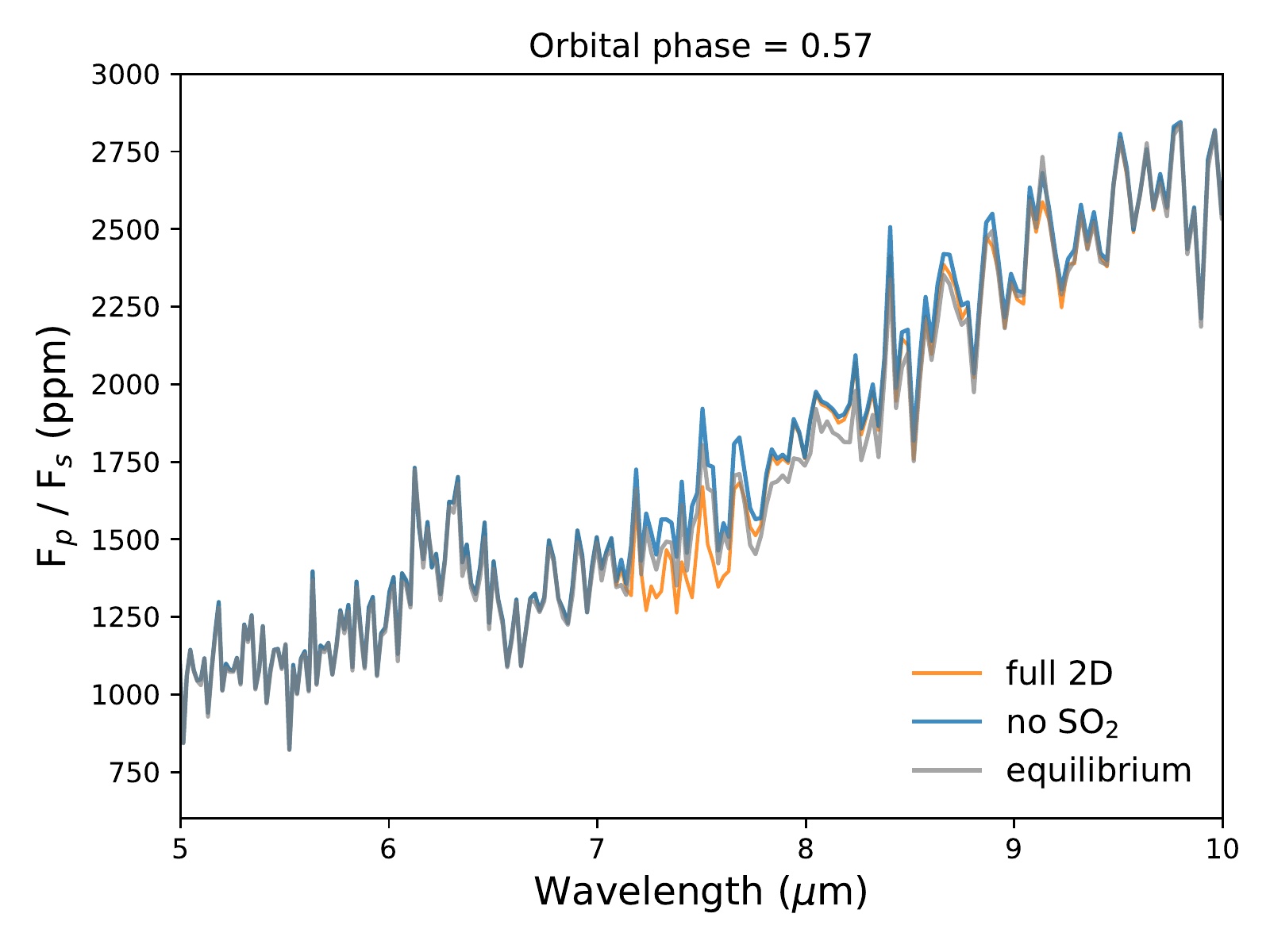}
\includegraphics[width=0.333\columnwidth]{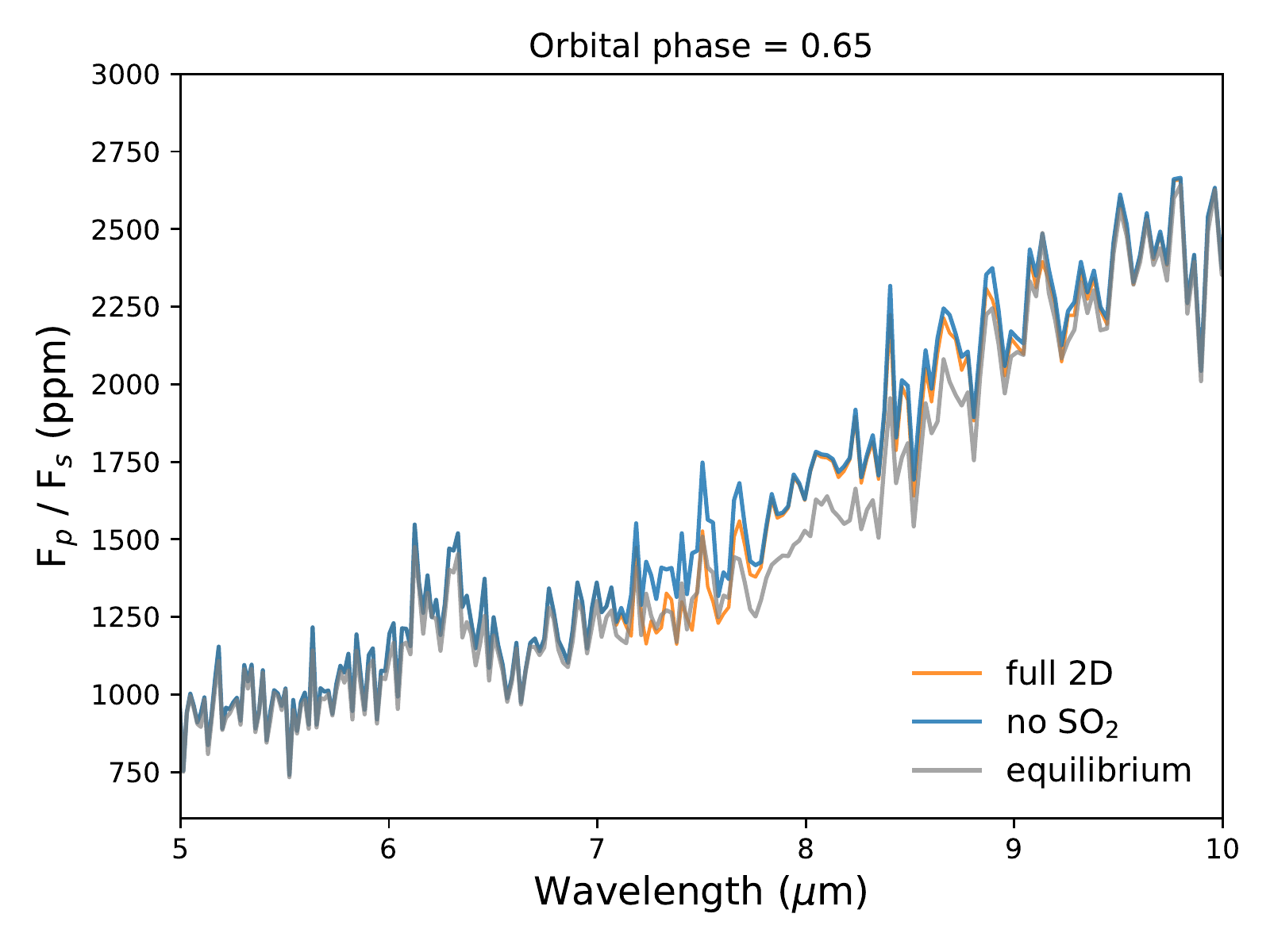}
\includegraphics[width=0.333\columnwidth]{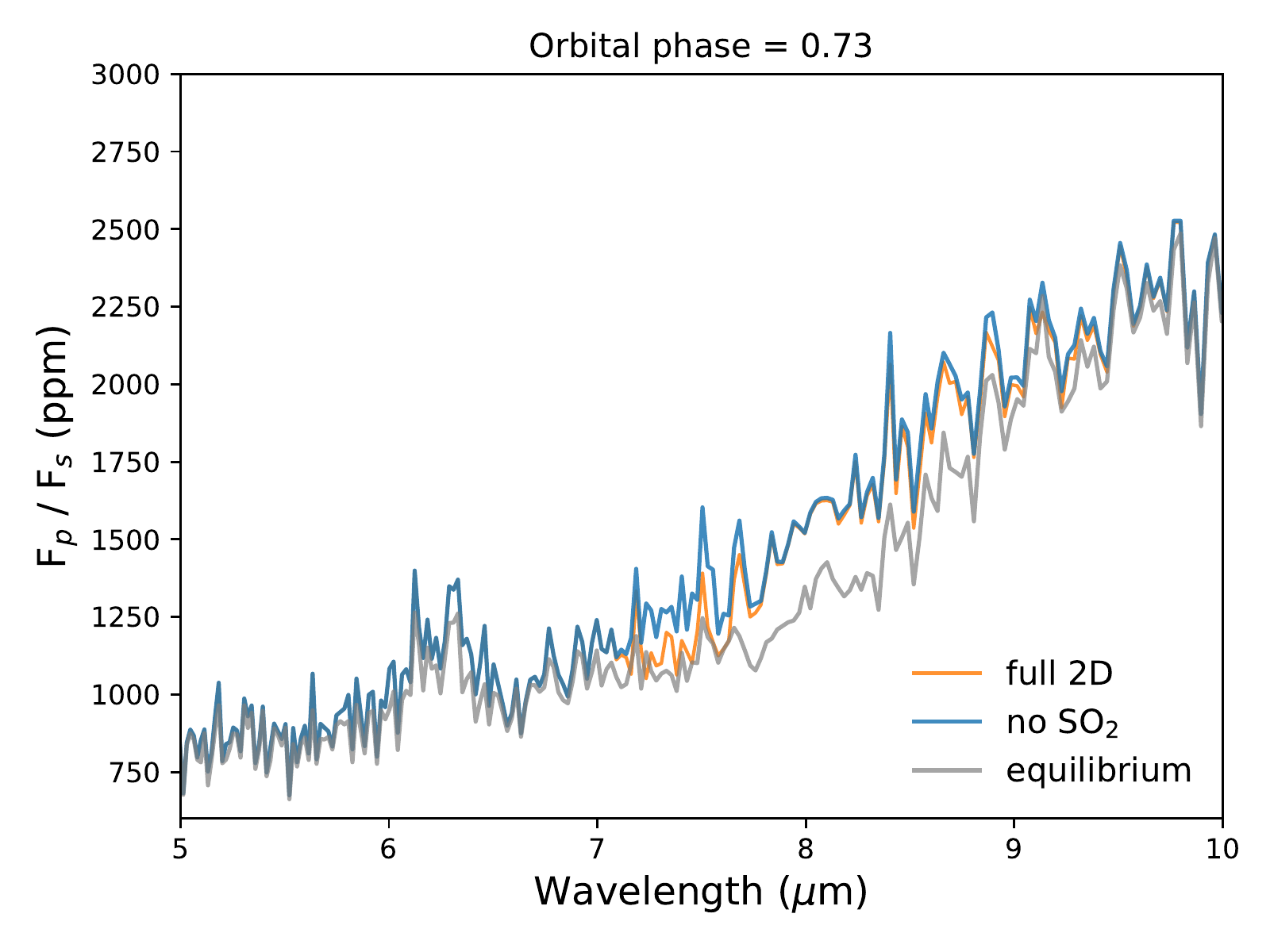}
\includegraphics[width=0.333\columnwidth]{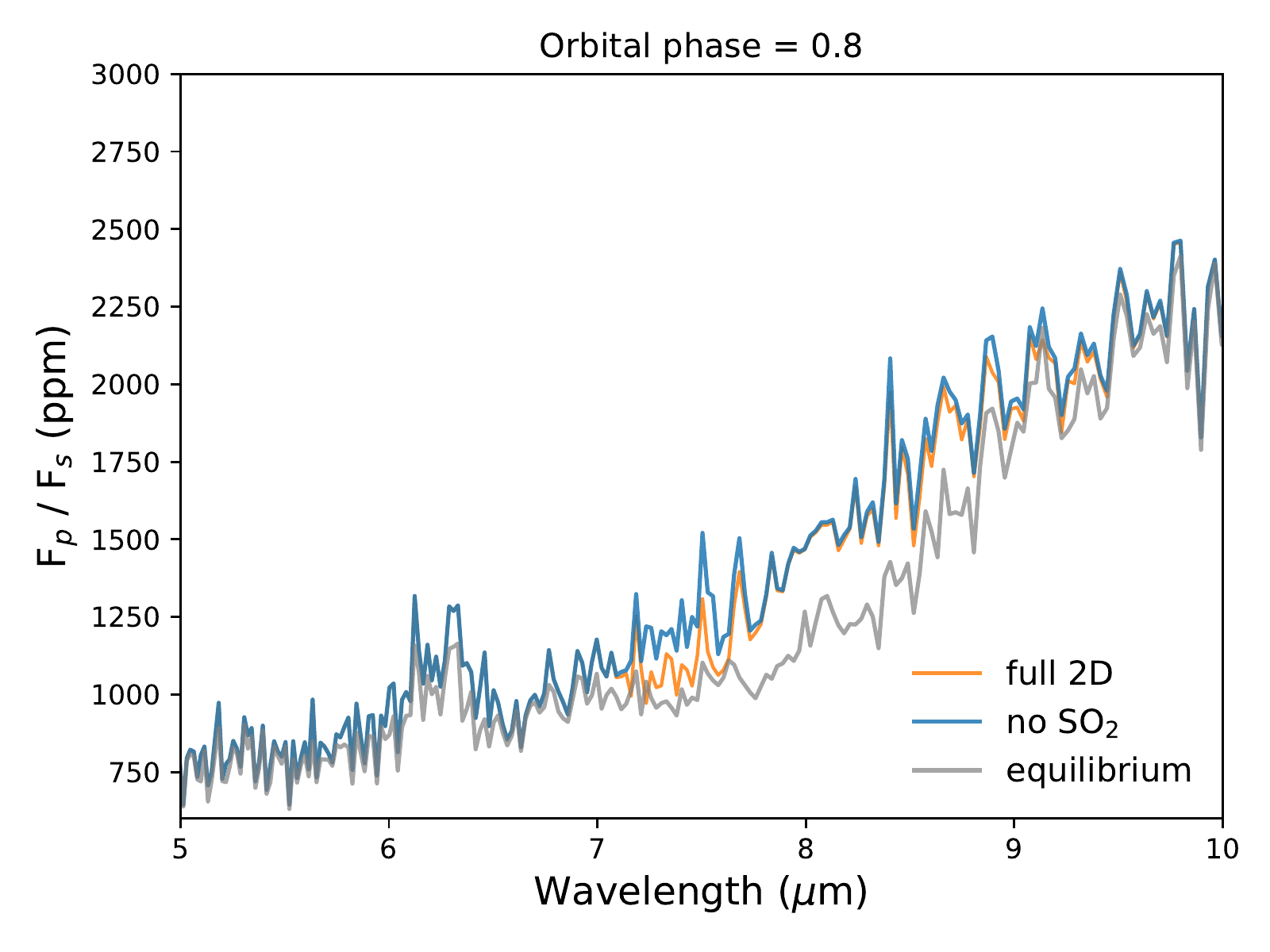}
\includegraphics[width=0.333\columnwidth]{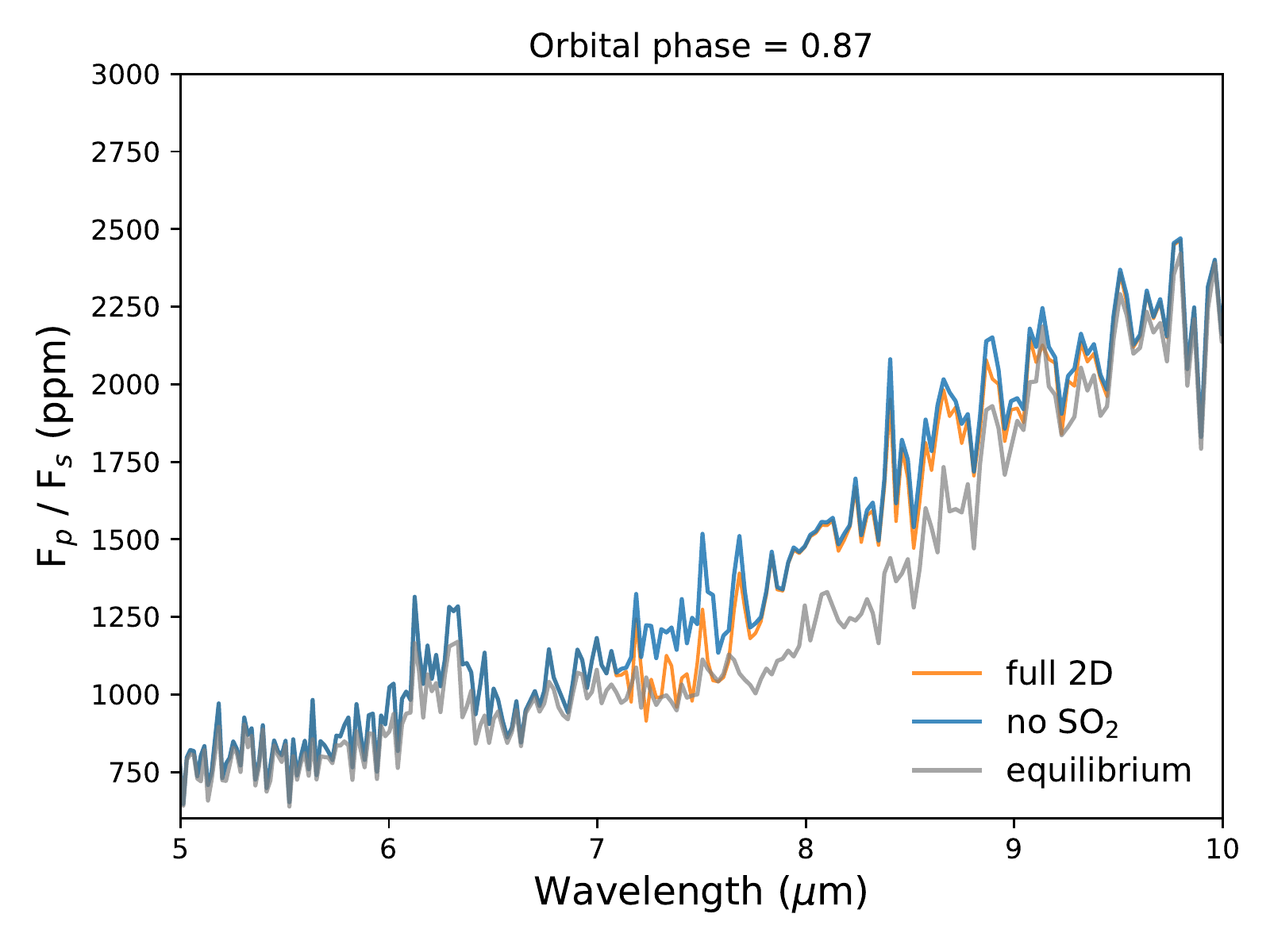}
\includegraphics[width=0.333\columnwidth]{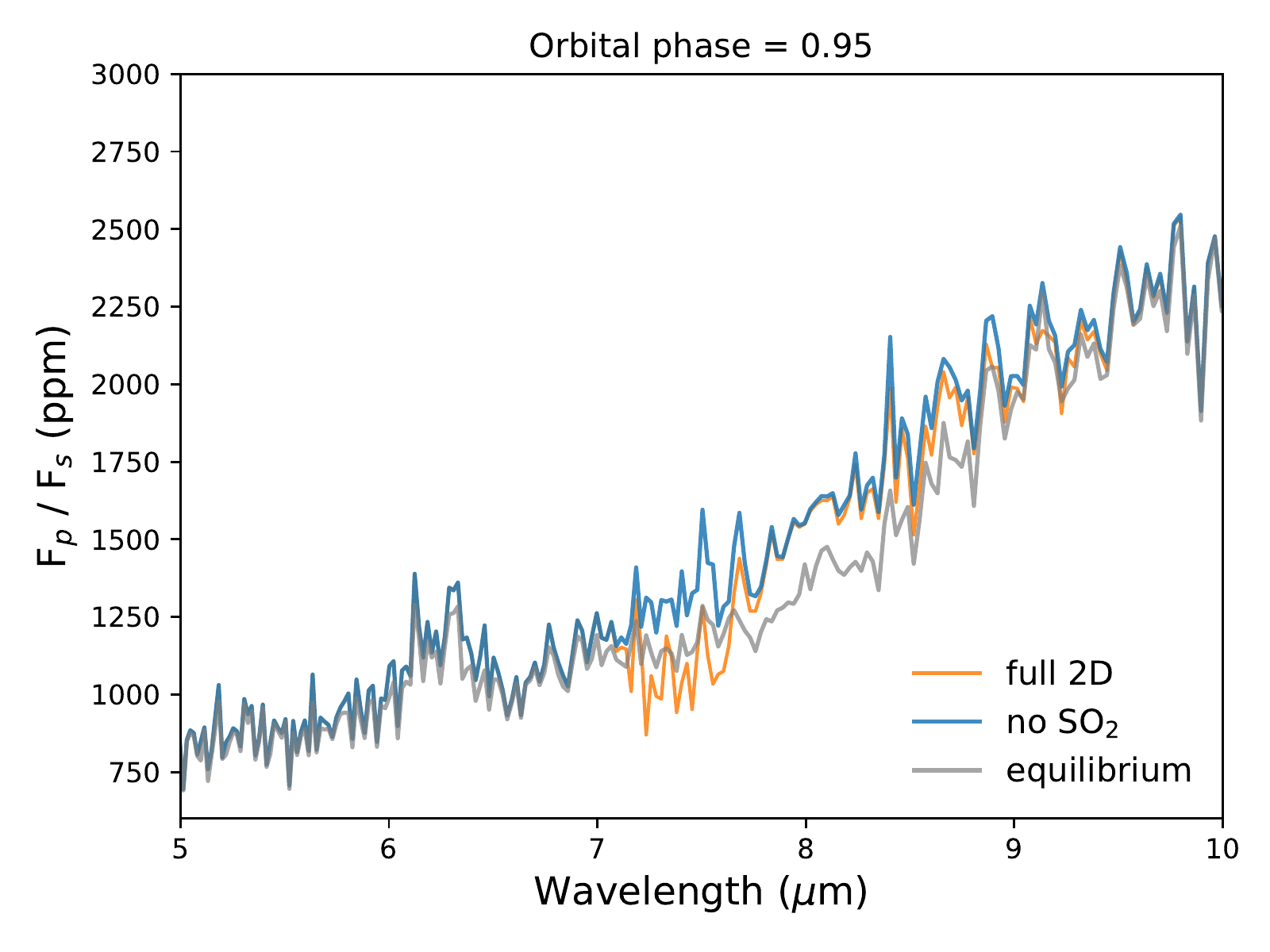}
\caption{Synthetic thermal emission spectra at different orbital phases (transit at 0 and eclipse at 0.5) with gas composition given by our 2D WASP-39b models (orange). The model without \ce{SO2} opacity is shown in blue and the model assuming chemical equilibrium is shown in gray for comparison.
}
\label{fig:emission} 
\end{figure}

The planetary emission spectra at different orbital phases are displayed in Figure \ref{fig:emission}. \ce{SO2} absorption between 7 and 8 $\mu m$ is consistently visible throughout the orbit. Notably, this feature becomes more prominent, reaching approximately 400 ppm, during the interval after transit and before the eclipse, i.e., when the nightside and evening hemisphere are in view. We find the orbital phase between 0 and 0.25 -- when the nightside and a portion of the evening side are facing Earth -- is likely to be the most favorable period for observing the transport of \ce{SO2} to the nightside hemisphere. Although \ce{SO2} has similar abundances on each limb, the more isothermal temperature on the morning limb reduces the absorption feature around the orbital phase 0.75. After the secondary eclipse, during the orbital phase between $\sim$ 0.6 and 0.9, the emission shows the greatest deviation from that predicted by chemical equilibrium, as \ce{CH4} remains deficient compared to its equilibrium abundance.



\section{Discussion}
The homogenizing effects of horizontal transport we demonstrated for WASP-39b can be extended to other synchronous rotating planets that feature fast zonal jets induced by day-night thermal forcing. As long as the timescale of horizontal transport is not overwhelmed by that of vertical mixing or very short chemical lifetimes, we expect a general occurrence of photochemical product redistribution towards the nightside. Other irradiated gas giants with less inflated scale height than WASP-39b are likely to have the upper atmosphere dominated by vertical mixing \citep[e.g.,][]{Agundez2014,Baeyens2021}. When vertical mixing outweighs horizontal transport in this region, the local interaction between photochemistry and vertical mixing is expected to enhance the compositional gradient in the zonal direction. In cases of non-synchronous or ultra-hot atmospheres that experience strong magnetic drag, the main circulation pattern deviates from equatorial superroation. Horizontal transport in such conditions can make distinct characteristics compared to the revolving transport by a zonal jet, which will be addressed in a separate paper (Tsai et al. submitted). Furthermore, although the same thermal and wind structures are used in this 2D study while exploring metallicity, we expect similar general transport properties with this small metallicity variation. However, the production of \ce{SO2} can cease and transition to elemental sulfur when the temperature drops below the transition point. This sensitivity to temperature can potentially affect the \ce{SO2} distribution and limb asymmetries. Future efforts to couple photochemistry with 3D general circulation models will provide a more complete understanding of global mixing \citep[e.g., ][]{Lee2023} and temporal variability \citep[e.g., ][]{Komacek2019}. 

Due to efficient heat redistribution, the nightside on WASP-39b is not cool enough for polysulfur to grow beyond \ce{S2}, nor for \ce{S8} to condense \citep[cf.][]{Zahnle2016,Tsai2023}. The laboratory experiments of \citet{reed20} suggest that organosulfur hazes can form in Archean-like atmospheres when \ce{CH4} and unsaturated hydrocarbon are present, but we find that the global abundances of \ce{CH4} and C$_2$H$_x$ hydrocarbons are significantly reduced when zonal winds are considered. It also remains to be investigated whether the formation of organosulfur hazes under a \ce{CO2}-rich condition \citep{He2020,Vuitton2021} can occur in a more reduced \ce{H2}-rich environment. On the other hand, we find that photochemically generated radicals and haze precursors can be effectively transported from their production regions on the dayside to the cooler nightside within the 2D modeling framework, which increases the likelihood of haze formation on the nightside. This becomes particularly relevant for exoplanets that are cooler than WASP-39b, such that \ce{CH4} is more likely present on the planet's dayside. Our modeling framework thus provides insight into how photochemical hazes can populate across the planet \citep[e.g.,][]{Kempton2017,Gao2020,Steinrueck2021}.

\begin{acknowledgments} 
The authors thank Taylor Bell for evaluating the phase curve observations. The authors thank Jeeyhun Yang and Sean Jordan for the discussions on the choice of reaction rate coefficients involving \ce{H2S}. S.-M.T. acknowledges support from NASA Exobiology Grant No. 80NSSC20K1437 and the University of California at Riverside. J.M. acknowledges support from NASA XRP grant number 80NSSC22K0314. D.P. acknowledges support from NASA through the NASA Hubble Fellowship grant HST-HF2-51490.001-A awarded by the Space Telescope Science Institute, which is operated by the Association of Universities for Research in Astronomy, Inc., for NASA, under contract NAS5-26555. E.K.H. Lee is supported by the SNSF Ambizione Fellowship grant (\#193448).
\end{acknowledgments}

%

\vspace{5mm}


\software{Exo-FMS \citep{Lee2020,Lee2021},  
          VULCAN \citep{tsai17,Tsai2021},
          Numpy \citep{numpy},
          Matplotlib \citep{matplotlib}
          }

\newpage
\appendix
\section{The lifetime of \ce{SO2}}
\begin{figure}[ht!]
\includegraphics[width=0.5\columnwidth]{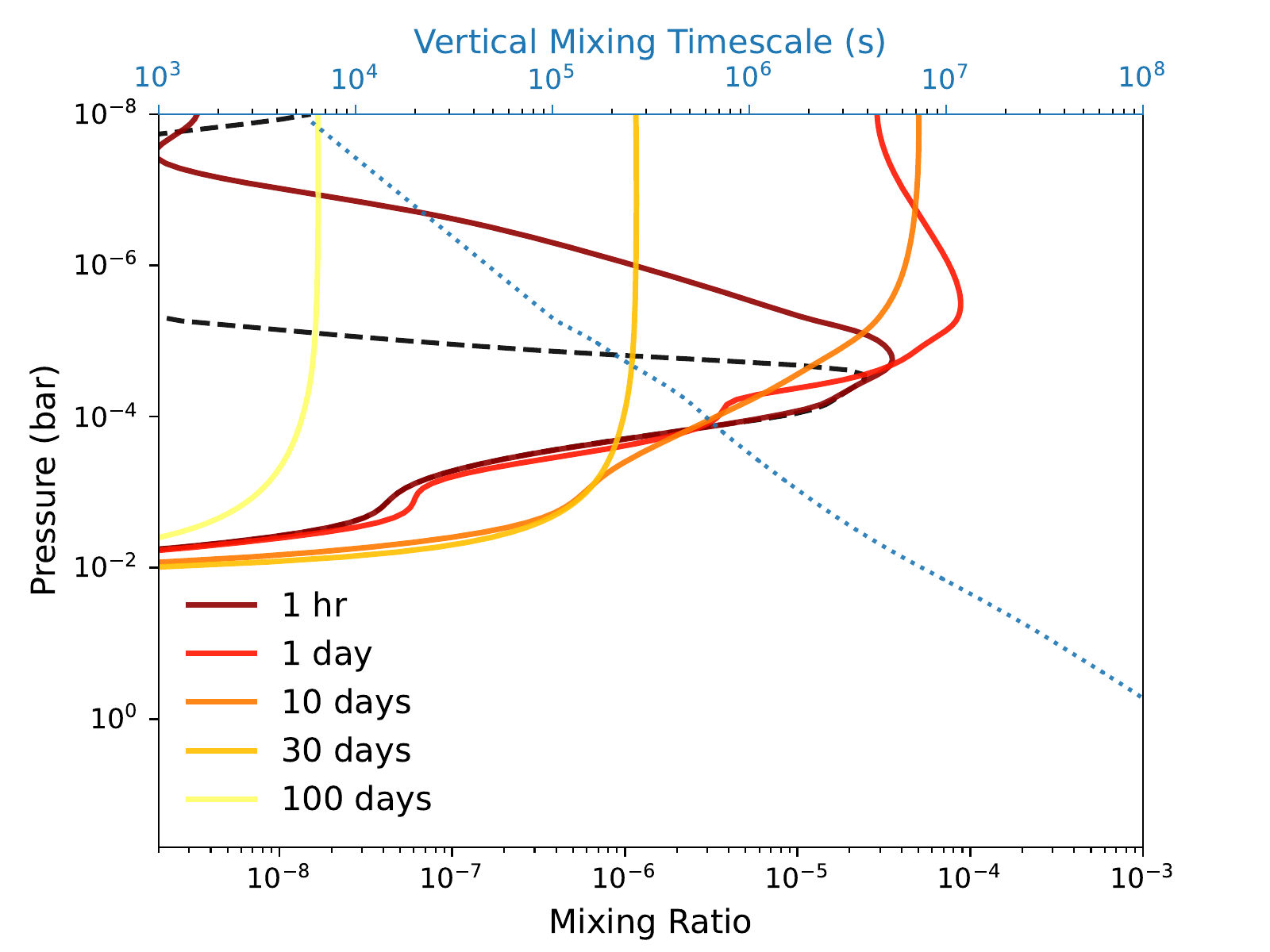}
\caption{The evolution of \ce{SO2} profile on the evening limb after the UV flux is switched off. The black dashed curve is the initial \ce{SO2} distribution, taken from the 1D VULCAN results in \cite{Tsai2023} (no horizontal transport from other longitudes) corresponding to 80$^{\circ}$ longitude. The blue dotted line shows the global vertical mixing timescale given by Equation (2) of \cite{Tsai2023}. This evolution demonstrates that \ce{SO2} would be dissipated by the vertical mixing process after $\gtrsim$ 10 days in the absence of stellar UV and horizontal transport.
} 
\label{fig:SO2_evo}
\end{figure}

\section{3D Exo-FMS GCM output}

\begin{table}[!h]
\centering
\caption{Exo-FMS GCM parameters for WASP-39b.}
\begin{tabular}{c c c l}  \hline \hline
 Symbol & Value  & Unit & Description \\ \hline
 T$_{\rm irr}$ & 1652 & K & Irradiation temperature \\
 T$_{\rm int}$ & 358 & K & Internal temperature \\
 P$_{\rm 0}$ & 220 &  bar & Lower boundary pressure \\
 P$_{\rm up}$ & 10$^{-6}$ &  bar & Upper boundary pressure \\
 c$_{\rm p}$ & 11335  &  J K$^{-1}$ kg$^{-1}$ & Specific heat capacity \\
 R$_{\rm d}$ & 3221 &  J K$^{-1}$ kg$^{-1}$  & Specific gas constant \\
 $\kappa$ &  0.284 & -  & Adiabatic lapse rate \\
 g$_{\rm p}$ & 4.26  & m s$^{-2}$ & Gravitational acceleration \\
 R$_{\rm p}$ & 9.14 $\cdot$ 10$^{7}$  & m & Planetary radius\\
 $\Omega_{\rm p}$ & 1.79 $\cdot$ 10$^{-5}$ & rad s$^{-1}$ & Planetary rotation rate \\
 N$_{\rm v}$ & 54  & - & Vertical resolution \\
 d$_{\rm 4}$ & 0.16  & - & $\mathcal{O}$(4) divergence dampening coefficient \\
\hline
\end{tabular}
\label{tab:GCM_parameters}
\end{table}

\begin{figure}[ht!]
\includegraphics[width=0.5\columnwidth]{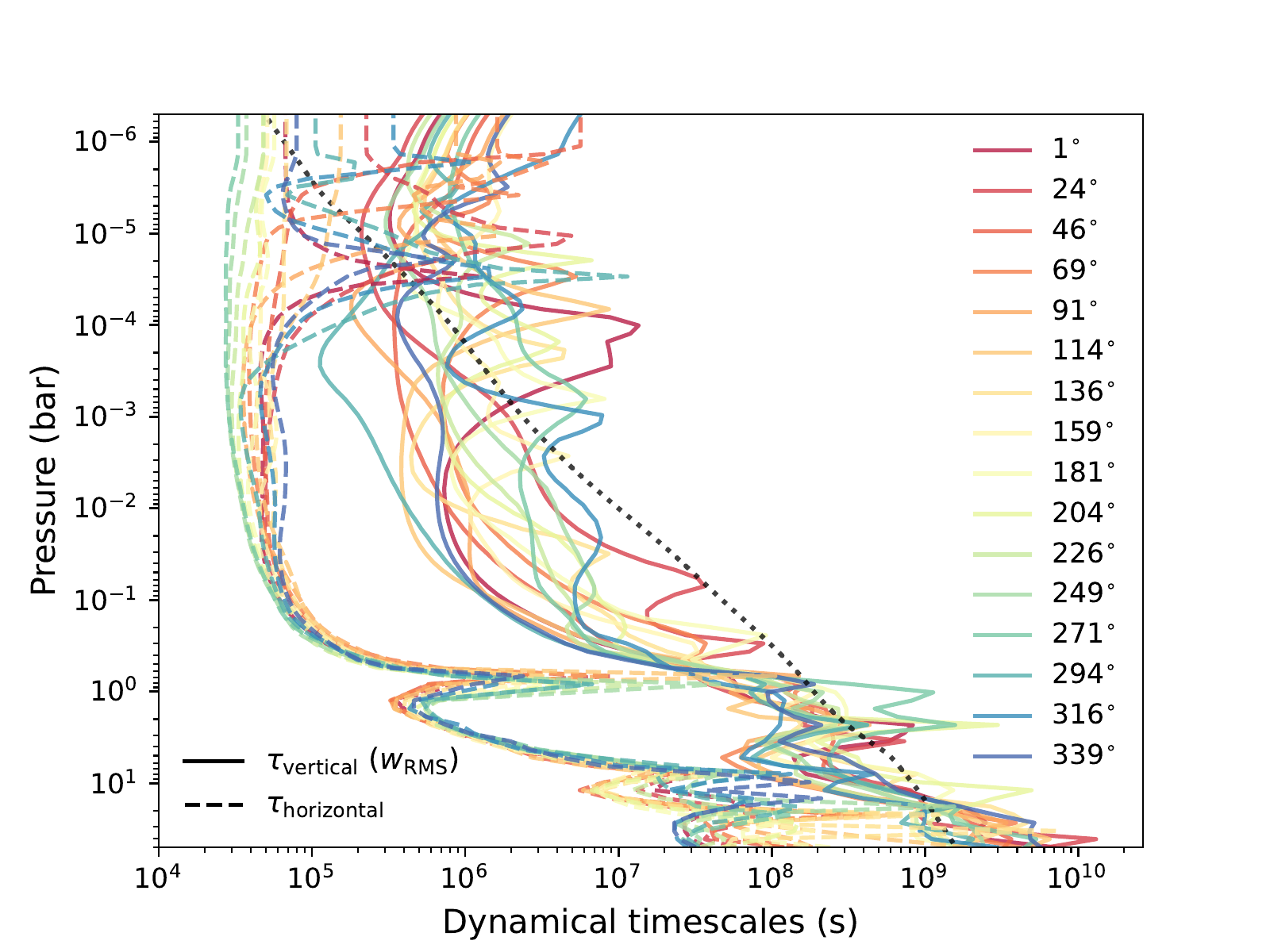}
\includegraphics[width=0.5\columnwidth]{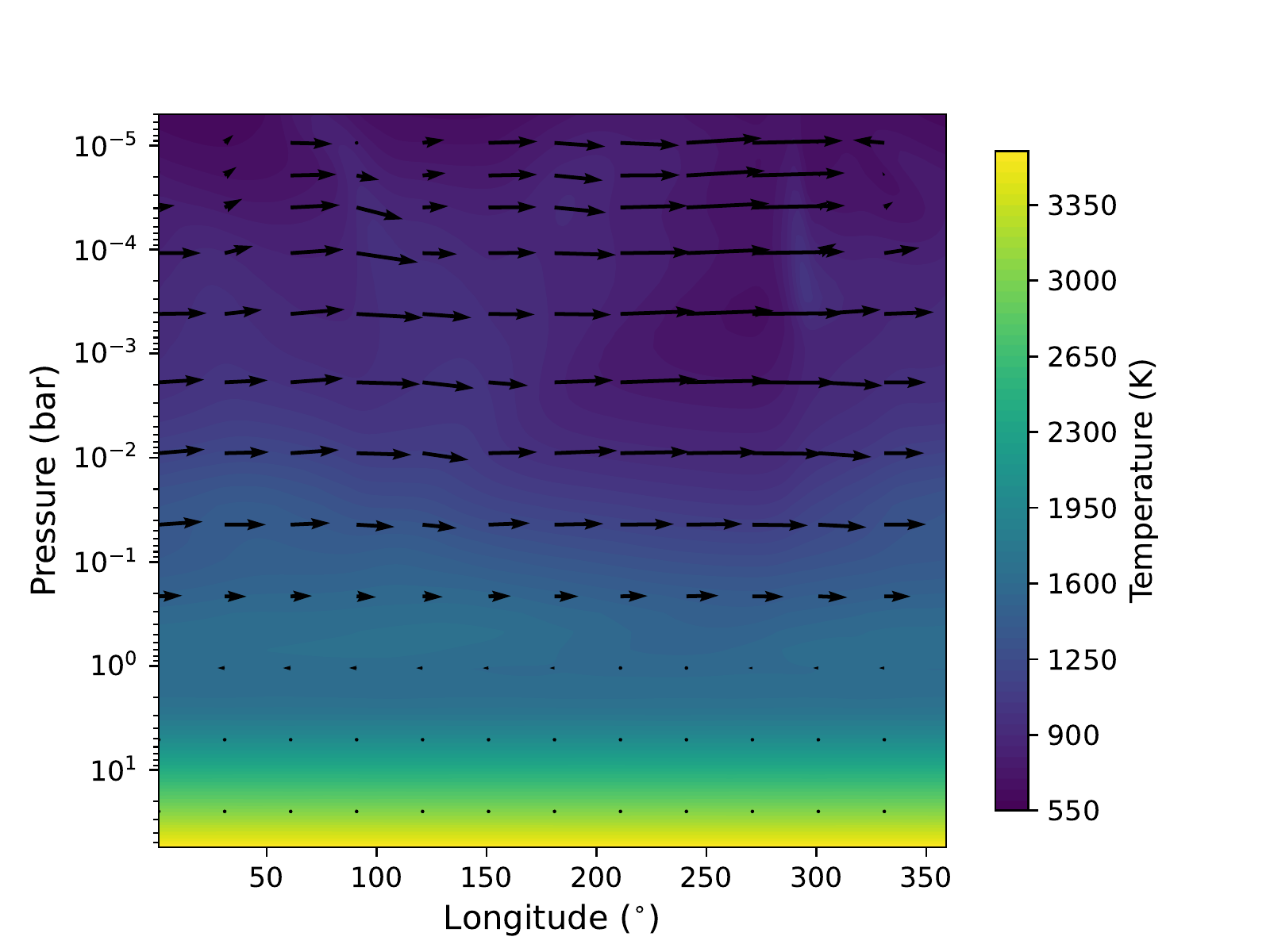}
\includegraphics[width=0.5\columnwidth]{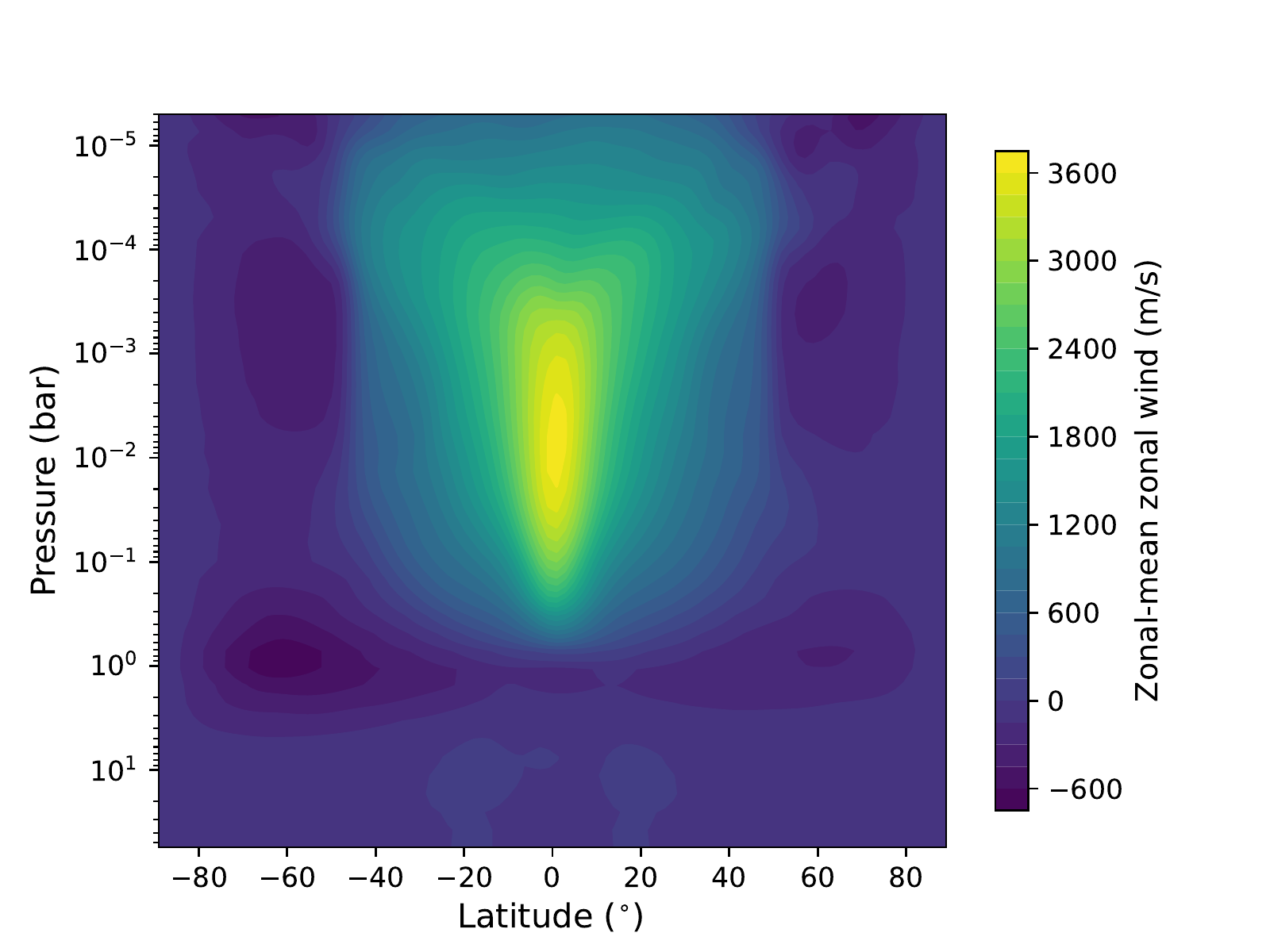}
\includegraphics[width=0.5\columnwidth]{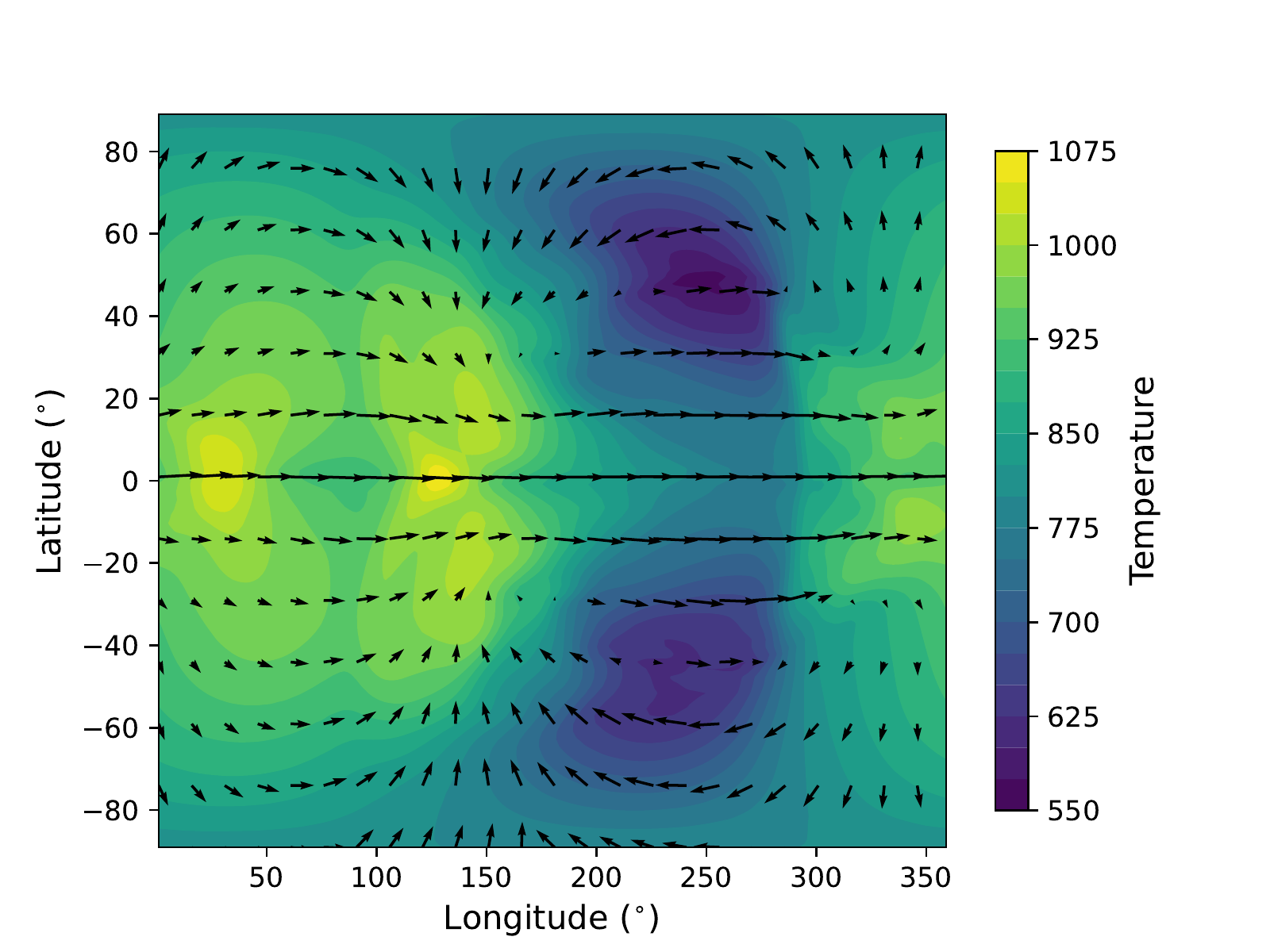}
\caption{The top left panel illustrates the dynamical timescales of vertical mixing (solid) and horizontal transport (dashed) derived from GCM winds. The vertical mixing timescale using the fitting function in \cite{Tsai2023} is shown in the black dot line. Horizontal transport dominates below 10$^{-4}$ bar. The top right panel shows the temperatures (color scale) and winds (arrows) on the meridionally averaged equatorial plane (substellar point located at 0$^{\circ}$ longitude). The aspect ratio of the wind vector has been re-scaled to 1:10 to exaggerate the vertical velocity component. The bottom left panel shows the zonal-mean zonal wind and the bottom right panel shows the temperatures and winds at approximately 1.3 mbar level.}
\label{fig:tau_dyn} 
\end{figure}

\bibliographystyle{aasjournal}
\bibliography{master_bib.bib}

\begin{thebibliography}{}
\expandafter\ifx\csname natexlab\endcsname\relax\def\natexlab#1{#1}\fi
\providecommand{\url}[1]{\href{#1}{#1}}
\providecommand{\dodoi}[1]{doi:~\href{http://doi.org/#1}{\nolinkurl{#1}}}
\providecommand{\doeprint}[1]{\href{http://ascl.net/#1}{\nolinkurl{http://ascl.net/#1}}}
\providecommand{\doarXiv}[1]{\href{https://arxiv.org/abs/#1}{\nolinkurl{https://arxiv.org/abs/#1}}}

\bibitem[{{Ag{\'u}ndez} {et~al.}(2014){Ag{\'u}ndez}, {Venot}, {Selsis}, \&
  {Iro}}]{Agundez2014}
{Ag{\'u}ndez}, M., {Venot}, O., {Selsis}, F., \& {Iro}, N. 2014, \apj, 781, 68,
  \dodoi{10.1088/0004-637X/781/2/68}

\bibitem[{{Ahrer} {et~al.}(2023){Ahrer}, {Stevenson}, {Mansfield}, {Moran},
  {Brande}, {Morello}, {Murray}, {Nikolov}, {Petit dit de la Roche},
  {Schlawin}, {Wheatley}, {Zieba}, {Batalha}, {Damiano}, {Goyal}, {Lendl},
  {Lothringer}, {Mukherjee}, {Ohno}, {Batalha}, {Battley}, {Bean}, {Beatty},
  {Benneke}, {Berta-Thompson}, {Carter}, {Cubillos}, {Daylan}, {Espinoza},
  {Gao}, {Gibson}, {Gill}, {Harrington}, {Hu}, {Kreidberg}, {Lewis}, {Line},
  {L{\'o}pez-Morales}, {Parmentier}, {Powell}, {Sing}, {Tsai}, {Wakeford},
  {Welbanks}, {Alam}, {Alderson}, {Allen}, {Anderson}, {Barstow}, {Bayliss},
  {Bell}, {Blecic}, {Bryant}, {Burleigh}, {Carone}, {Casewell}, {Changeat},
  {Chubb}, {Crossfield}, {Crouzet}, {Decin}, {D{\'e}sert}, {Feinstein},
  {Flagg}, {Fortney}, {Gizis}, {Heng}, {Iro}, {Kempton}, {Kendrew}, {Kirk},
  {Knutson}, {Komacek}, {Lagage}, {Leconte}, {Lustig-Yaeger}, {MacDonald},
  {Mancini}, {May}, {Mayne}, {Miguel}, {Mikal-Evans}, {Molaverdikhani},
  {Palle}, {Piaulet}, {Rackham}, {Redfield}, {Rogers}, {Roy}, {Rustamkulov},
  {Shkolnik}, {Sotzen}, {Taylor}, {Tremblin}, {Tucker}, {Turner}, {de
  Val-Borro}, {Venot}, \& {Zhang}}]{Ahrer2023}
{Ahrer}, E.-M., {Stevenson}, K.~B., {Mansfield}, M., {et~al.} 2023, \nat, 614,
  653, \dodoi{10.1038/s41586-022-05590-4}

\bibitem[{{Alderson} {et~al.}(2023){Alderson}, {Wakeford}, {Alam}, {Batalha},
  {Lothringer}, {Adams Redai}, {Barat}, {Brande}, {Damiano}, {Daylan},
  {Espinoza}, {Flagg}, {Goyal}, {Grant}, {Hu}, {Inglis}, {Lee}, {Mikal-Evans},
  {Ramos-Rosado}, {Roy}, {Wallack}, {Batalha}, {Bean}, {Benneke},
  {Berta-Thompson}, {Carter}, {Changeat}, {Col{\'o}n}, {Crossfield},
  {D{\'e}sert}, {Foreman-Mackey}, {Gibson}, {Kreidberg}, {Line},
  {L{\'o}pez-Morales}, {Molaverdikhani}, {Moran}, {Morello}, {Moses},
  {Mukherjee}, {Schlawin}, {Sing}, {Stevenson}, {Taylor}, {Aggarwal}, {Ahrer},
  {Allen}, {Barstow}, {Bell}, {Blecic}, {Casewell}, {Chubb}, {Crouzet},
  {Cubillos}, {Decin}, {Feinstein}, {Fortney}, {Harrington}, {Heng}, {Iro},
  {Kempton}, {Kirk}, {Knutson}, {Krick}, {Leconte}, {Lendl}, {MacDonald},
  {Mancini}, {Mansfield}, {May}, {Mayne}, {Miguel}, {Nikolov}, {Ohno}, {Palle},
  {Parmentier}, {Petit dit de la Roche}, {Piaulet}, {Powell}, {Rackham},
  {Redfield}, {Rogers}, {Rustamkulov}, {Tan}, {Tremblin}, {Tsai}, {Turner}, {de
  Val-Borro}, {Venot}, {Welbanks}, {Wheatley}, \& {Zhang}}]{Alderson2023}
{Alderson}, L., {Wakeford}, H.~R., {Alam}, M.~K., {et~al.} 2023, \nat, 614,
  664, \dodoi{10.1038/s41586-022-05591-3}

\bibitem[{{Baeyens} {et~al.}(2021){Baeyens}, {Decin}, {Carone}, {Venot},
  {Ag{\'u}ndez}, \& {Molli{\`e}re}}]{Baeyens2021}
{Baeyens}, R., {Decin}, L., {Carone}, L., {et~al.} 2021, \mnras, 505, 5603,
  \dodoi{10.1093/mnras/stab1310}

\bibitem[{{Baeyens} {et~al.}(2022){Baeyens}, {Konings}, {Venot}, {Carone}, \&
  {Decin}}]{Baeyens2022}
{Baeyens}, R., {Konings}, T., {Venot}, O., {Carone}, L., \& {Decin}, L. 2022,
  \mnras, 512, 4877, \dodoi{10.1093/mnras/stac809}

\bibitem[{{Chen} {et~al.}(2021){Chen}, {Zhan}, {Youngblood}, {Wolf},
  {Feinstein}, \& {Horton}}]{Chen2021}
{Chen}, H., {Zhan}, Z., {Youngblood}, A., {et~al.} 2021, Nature Astronomy, 5,
  298, \dodoi{10.1038/s41550-020-01264-1}

\bibitem[{{Chubb} \& {Min}(2022)}]{Chubb2022}
{Chubb}, K.~L., \& {Min}, M. 2022, \aap, 665, A2,
  \dodoi{10.1051/0004-6361/202142800}

\bibitem[{{Cooper} \& {Showman}(2006)}]{Cooper2006}
{Cooper}, C.~S., \& {Showman}, A.~P. 2006, \apj, 649, 1048,
  \dodoi{10.1086/506312}

\bibitem[{{Crossfield}(2023)}]{Crossfield2023}
{Crossfield}, I. J.~M. 2023, arXiv e-prints, arXiv:2303.17622,
  \dodoi{10.48550/arXiv.2303.17622}

\bibitem[{{Dobbs-Dixon} {et~al.}(2012){Dobbs-Dixon}, {Agol}, \&
  {Burrows}}]{Ian2012}
{Dobbs-Dixon}, I., {Agol}, E., \& {Burrows}, A. 2012, \apj, 751, 87,
  \dodoi{10.1088/0004-637X/751/2/87}

\bibitem[{{Drummond} {et~al.}(2020){Drummond}, {H{\'e}brard}, {Mayne}, {Venot},
  {Ridgway}, {Changeat}, {Tsai}, {Manners}, {Tremblin}, {Abraham}, {Sing}, \&
  {Kohary}}]{Drummond2020}
{Drummond}, B., {H{\'e}brard}, E., {Mayne}, N.~J., {et~al.} 2020, \aap, 636,
  A68, \dodoi{10.1051/0004-6361/201937153}

\bibitem[{{Espinoza} \& {Jones}(2021)}]{Espinoza2021}
{Espinoza}, N., \& {Jones}, K. 2021, \aj, 162, 165,
  \dodoi{10.3847/1538-3881/ac134d}

\bibitem[{{Feinstein} {et~al.}(2023){Feinstein}, {Radica}, {Welbanks},
  {Murray}, {Ohno}, {Coulombe}, {Espinoza}, {Bean}, {Teske}, {Benneke}, {Line},
  {Rustamkulov}, {Saba}, {Tsiaras}, {Barstow}, {Fortney}, {Gao}, {Knutson},
  {MacDonald}, {Mikal-Evans}, {Rackham}, {Taylor}, {Parmentier}, {Batalha},
  {Berta-Thompson}, {Carter}, {Changeat}, {dos Santos}, {Gibson}, {Goyal},
  {Kreidberg}, {L{\'o}pez-Morales}, {Lothringer}, {Miguel}, {Molaverdikhani},
  {Moran}, {Morello}, {Mukherjee}, {Sing}, {Stevenson}, {Wakeford}, {Ahrer},
  {Alam}, {Alderson}, {Allen}, {Batalha}, {Bell}, {Blecic}, {Brande},
  {Caceres}, {Casewell}, {Chubb}, {Crossfield}, {Crouzet}, {Cubillos}, {Decin},
  {D{\'e}sert}, {Harrington}, {Heng}, {Henning}, {Iro}, {Kempton}, {Kendrew},
  {Kirk}, {Krick}, {Lagage}, {Lendl}, {Mancini}, {Mansfield}, {May}, {Mayne},
  {Nikolov}, {Palle}, {Petit dit de la Roche}, {Piaulet}, {Powell}, {Redfield},
  {Rogers}, {Roman}, {Roy}, {Nixon}, {Schlawin}, {Tan}, {Tremblin}, {Turner},
  {Venot}, {Waalkes}, {Wheatley}, \& {Zhang}}]{Feinstein2023}
{Feinstein}, A.~D., {Radica}, M., {Welbanks}, L., {et~al.} 2023, \nat, 614,
  670, \dodoi{10.1038/s41586-022-05674-1}

\bibitem[{{Gao} {et~al.}(2020){Gao}, {Thorngren}, {Lee}, {Fortney}, {Morley},
  {Wakeford}, {Powell}, {Stevenson}, \& {Zhang}}]{Gao2020}
{Gao}, P., {Thorngren}, D.~P., {Lee}, E. K.~H., {et~al.} 2020, Nature
  Astronomy, 4, 951, \dodoi{10.1038/s41550-020-1114-3}

\bibitem[{{Grant} \& {Wakeford}(2023)}]{Grant2023}
{Grant}, D., \& {Wakeford}, H.~R. 2023, \mnras, 519, 5114,
  \dodoi{10.1093/mnras/stac3632}

\bibitem[{He {et~al.}(2020)He, Hörst, Lewis, Yu, Moses, McGuiggan, Marley,
  Kempton, Morley, Valenti, \& Vuitton}]{He2020}
He, C., Hörst, S.~M., Lewis, N.~K., {et~al.} 2020, The Planetary Science
  Journal, 1, 51, \dodoi{10.3847/psj/abb1a4}

\bibitem[{Hunter(2007)}]{matplotlib}
Hunter, J.~D. 2007, Computing in Science Engineering, 9, 90,
  \dodoi{10.1109/MCSE.2007.55}

\bibitem[{{JWST Transiting Exoplanet Community Early Release Science Team}
  {et~al.}(2023){JWST Transiting Exoplanet Community Early Release Science
  Team}, {Ahrer}, {Alderson}, {Batalha}, {Batalha}, {Bean}, {Beatty}, {Bell},
  {Benneke}, {Berta-Thompson}, {Carter}, {Crossfield}, {Espinoza}, {Feinstein},
  {Fortney}, {Gibson}, {Goyal}, {Kempton}, {Kirk}, {Kreidberg},
  {L{\'o}pez-Morales}, {Line}, {Lothringer}, {Moran}, {Mukherjee}, {Ohno},
  {Parmentier}, {Piaulet}, {Rustamkulov}, {Schlawin}, {Sing}, {Stevenson},
  {Wakeford}, {Allen}, {Birkmann}, {Brande}, {Crouzet}, {Cubillos}, {Damiano},
  {D{\'e}sert}, {Gao}, {Harrington}, {Hu}, {Kendrew}, {Knutson}, {Lagage},
  {Leconte}, {Lendl}, {MacDonald}, {May}, {Miguel}, {Molaverdikhani}, {Moses},
  {Murray}, {Nehring}, {Nikolov}, {Petit dit de la Roche}, {Radica}, {Roy},
  {Stassun}, {Taylor}, {Waalkes}, {Wachiraphan}, {Welbanks}, {Wheatley},
  {Aggarwal}, {Alam}, {Banerjee}, {Barstow}, {Blecic}, {Casewell}, {Changeat},
  {Chubb}, {Col{\'o}n}, {Coulombe}, {Daylan}, {de Val-Borro}, {Decin}, {Dos
  Santos}, {Flagg}, {France}, {Fu}, {Garc{\'\i}a Mu{\~n}oz}, {Gizis},
  {Glidden}, {Grant}, {Heng}, {Henning}, {Hong}, {Inglis}, {Iro}, {Kataria},
  {Komacek}, {Krick}, {Lee}, {Lewis}, {Lillo-Box}, {Lustig-Yaeger}, {Mancini},
  {Mandell}, {Mansfield}, {Marley}, {Mikal-Evans}, {Morello}, {Nixon}, {Ortiz
  Ceballos}, {Piette}, {Powell}, {Rackham}, {Ramos-Rosado}, {Rauscher},
  {Redfield}, {Rogers}, {Roman}, {Roudier}, {Scarsdale}, {Shkolnik},
  {Southworth}, {Spake}, {Steinrueck}, {Tan}, {Teske}, {Tremblin}, {Tsai},
  {Tucker}, {Turner}, {Valenti}, {Venot}, {Waldmann}, {Wallack}, {Zhang}, \&
  {Zieba}}]{ERS2023}
{JWST Transiting Exoplanet Community Early Release Science Team}, {Ahrer},
  E.-M., {Alderson}, L., {et~al.} 2023, \nat, 614, 649,
  \dodoi{10.1038/s41586-022-05269-w}

\bibitem[{{Kempton} {et~al.}(2017){Kempton}, {Bean}, \&
  {Parmentier}}]{Kempton2017}
{Kempton}, E. M.~R., {Bean}, J.~L., \& {Parmentier}, V. 2017, \apjl, 845, L20,
  \dodoi{10.3847/2041-8213/aa84ac}

\bibitem[{{Kempton} {et~al.}(2023){Kempton}, {Zhang}, {Bean}, {Steinrueck},
  {Piette}, {Parmentier}, {Malsky}, {Roman}, {Rauscher}, {Gao}, {Bell}, {Xue},
  {Taylor}, {Savel}, {Arnold}, {Nixon}, {Stevenson}, {Mansfield}, {Kendrew},
  {Zieba}, {Ducrot}, {Dyrek}, {Lagage}, {Stassun}, {Henry}, {Barman}, {Lupu},
  {Malik}, {Kataria}, {Ih}, {Fu}, {Welbanks}, \& {McGill}}]{Kempton2023}
{Kempton}, E. M.~R., {Zhang}, M., {Bean}, J.~L., {et~al.} 2023, arXiv e-prints,
  arXiv:2305.06240, \dodoi{10.48550/arXiv.2305.06240}

\bibitem[{{Komacek} {et~al.}(2019){Komacek}, {Showman}, \&
  {Parmentier}}]{Komacek2019}
{Komacek}, T.~D., {Showman}, A.~P., \& {Parmentier}, V. 2019, \apj, 881, 152,
  \dodoi{10.3847/1538-4357/ab338b}

\bibitem[{{Lee} {et~al.}(2016){Lee}, {Dobbs-Dixon}, {Helling}, {Bognar}, \&
  {Woitke}}]{Lee2016}
{Lee}, E., {Dobbs-Dixon}, I., {Helling}, C., {Bognar}, K., \& {Woitke}, P.
  2016, \aap, 594, A48, \dodoi{10.1051/0004-6361/201628606}

\bibitem[{Lee {et~al.}(2020)Lee, Casewell, Chubb, Hammond, Tan, Tsai, \&
  Pierrehumbert}]{Lee2020}
Lee, E.~K., Casewell, S.~L., Chubb, K.~L., {et~al.} 2020, Monthly Notices of
  the Royal Astronomical Society, 496, 4674, \dodoi{10.1093/mnras/staa1882}

\bibitem[{Lee {et~al.}(2021)Lee, Parmentier, Hammond, Grimm, Kitzmann, Tan,
  Tsai, \& Pierrehumbert}]{Lee2021}
Lee, E. K.~H., Parmentier, V., Hammond, M., {et~al.} 2021, MNRAS, 506, 2695.
\newblock \doarXiv{2106.11664}

\bibitem[{{Lee} {et~al.}(2023){Lee}, {Tsai}, {Hammond}, \& {Tan}}]{Lee2023}
{Lee}, E. K.~H., {Tsai}, S.-M., {Hammond}, M., \& {Tan}, X. 2023, \aap, 672,
  A110, \dodoi{10.1051/0004-6361/202245473}

\bibitem[{{Line} {et~al.}(2016){Line}, {Stevenson}, {Bean}, {Desert},
  {Fortney}, {Kreidberg}, {Madhusudhan}, {Showman}, \&
  {Diamond-Lowe}}]{Line2016}
{Line}, M.~R., {Stevenson}, K.~B., {Bean}, J., {et~al.} 2016, \aj, 152, 203,
  \dodoi{10.3847/0004-6256/152/6/203}

\bibitem[{{Lines} {et~al.}(2018){Lines}, {Mayne}, {Boutle}, {Manners}, {Lee},
  {Helling}, {Drummond}, {Amundsen}, {Goyal}, {Acreman}, {Tremblin}, \&
  {Kerslake}}]{Lines2018}
{Lines}, S., {Mayne}, N.~J., {Boutle}, I.~A., {et~al.} 2018, \aap, 615, A97,
  \dodoi{10.1051/0004-6361/201732278}

\bibitem[{{MacDonald} {et~al.}(2020){MacDonald}, {Goyal}, \&
  {Lewis}}]{MacDonald2020}
{MacDonald}, R.~J., {Goyal}, J.~M., \& {Lewis}, N.~K. 2020, \apjl, 893, L43,
  \dodoi{10.3847/2041-8213/ab8238}

\bibitem[{Malik {et~al.}(2019)Malik, Kitzmann, Mendon{\c{c}}a, Grimm, Marleau,
  Linder, Tsai, \& Heng}]{Malik2019}
Malik, M., Kitzmann, D., Mendon{\c{c}}a, J.~M., {et~al.} 2019, The Astronomical
  Journal, 157, 170, \dodoi{10.3847/1538-3881/ab1084}

\bibitem[{{Mendon{\c{c}}a} {et~al.}(2018{\natexlab{a}}){Mendon{\c{c}}a},
  {Malik}, {Demory}, \& {Heng}}]{Mendonca2018}
{Mendon{\c{c}}a}, J.~M., {Malik}, M., {Demory}, B.-O., \& {Heng}, K.
  2018{\natexlab{a}}, \aj, 155, 150, \dodoi{10.3847/1538-3881/aaaebc}

\bibitem[{{Mendon{\c{c}}a} {et~al.}(2018{\natexlab{b}}){Mendon{\c{c}}a},
  {Tsai}, {Malik}, {Grimm}, \& {Heng}}]{Mendonca2018a}
{Mendon{\c{c}}a}, J.~M., {Tsai}, S.-m., {Malik}, M., {Grimm}, S.~L., \& {Heng},
  K. 2018{\natexlab{b}}, \apj, 869, 107, \dodoi{10.3847/1538-4357/aaed23}

\bibitem[{{Moses} {et~al.}(2021){Moses}, {Tremblin}, {Venot}, \&
  {Miguel}}]{Moses2021}
{Moses}, J.~I., {Tremblin}, P., {Venot}, O., \& {Miguel}, Y. 2021, Experimental
  Astronomy, \dodoi{10.1007/s10686-021-09749-1}

\bibitem[{{Parmentier} {et~al.}(2013){Parmentier}, {Showman}, \&
  {Lian}}]{Parmentier2013}
{Parmentier}, V., {Showman}, A.~P., \& {Lian}, Y. 2013, \aap, 558, A91,
  \dodoi{10.1051/0004-6361/201321132}

\bibitem[{{Reed} {et~al.}(2020){Reed}, {Browne}, \& {Tolbert}}]{reed20}
{Reed}, N.~W., {Browne}, E.~C., \& {Tolbert}, M.~A. 2020, ACS Earth and Space
  Chemistry, 4, 897, \dodoi{10.1021/acsearthspacechem.0c00086}

\bibitem[{{Rustamkulov} {et~al.}(2023){Rustamkulov}, {Sing}, {Mukherjee},
  {May}, {Kirk}, {Schlawin}, {Line}, {Piaulet}, {Carter}, {Batalha}, {Goyal},
  {L{\'o}pez-Morales}, {Lothringer}, {MacDonald}, {Moran}, {Stevenson},
  {Wakeford}, {Espinoza}, {Bean}, {Batalha}, {Benneke}, {Berta-Thompson},
  {Crossfield}, {Gao}, {Kreidberg}, {Powell}, {Cubillos}, {Gibson}, {Leconte},
  {Molaverdikhani}, {Nikolov}, {Parmentier}, {Roy}, {Taylor}, {Turner},
  {Wheatley}, {Aggarwal}, {Ahrer}, {Alam}, {Alderson}, {Allen}, {Banerjee},
  {Barat}, {Barrado}, {Barstow}, {Bell}, {Blecic}, {Brande}, {Casewell},
  {Changeat}, {Chubb}, {Crouzet}, {Daylan}, {Decin}, {D{\'e}sert},
  {Mikal-Evans}, {Feinstein}, {Flagg}, {Fortney}, {Harrington}, {Heng}, {Hong},
  {Hu}, {Iro}, {Kataria}, {Kempton}, {Krick}, {Lendl}, {Lillo-Box}, {Louca},
  {Lustig-Yaeger}, {Mancini}, {Mansfield}, {Mayne}, {Miguel}, {Morello},
  {Ohno}, {Palle}, {Petit dit de la Roche}, {Rackham}, {Radica},
  {Ramos-Rosado}, {Redfield}, {Rogers}, {Shkolnik}, {Southworth}, {Teske},
  {Tremblin}, {Tucker}, {Venot}, {Waalkes}, {Welbanks}, {Zhang}, \&
  {Zieba}}]{Zafra2023}
{Rustamkulov}, Z., {Sing}, D.~K., {Mukherjee}, S., {et~al.} 2023, \nat, 614,
  659, \dodoi{10.1038/s41586-022-05677-y}

\bibitem[{{Showman} {et~al.}(2020){Showman}, {Tan}, \&
  {Parmentier}}]{Showman2020}
{Showman}, A.~P., {Tan}, X., \& {Parmentier}, V. 2020, \ssr, 216, 139,
  \dodoi{10.1007/s11214-020-00758-8}

\bibitem[{Smith(1998)}]{Smith1998}
Smith, M.~D. 1998, Icarus, 132, 176 ,
  \dodoi{https://doi.org/10.1006/icar.1997.5886}

\bibitem[{Steinrueck {et~al.}(2021)Steinrueck, Showman, Lavvas, Koskinen, Tan,
  \& Zhang}]{Steinrueck2021}
Steinrueck, M.~E., Showman, A.~P., Lavvas, P., {et~al.} 2021, Monthly Notices
  of the Royal Astronomical Society, 504, 2783, \dodoi{10.1093/mnras/stab1053}

\bibitem[{{Stevenson} {et~al.}(2014){Stevenson}, {D{\'e}sert}, {Line}, {Bean},
  {Fortney}, {Showman}, {Kataria}, {Kreidberg}, {McCullough}, {Henry},
  {Charbonneau}, {Burrows}, {Seager}, {Madhusudhan}, {Williamson}, \&
  {Homeier}}]{Stevenson2014}
{Stevenson}, K.~B., {D{\'e}sert}, J.-M., {Line}, M.~R., {et~al.} 2014, Science,
  346, 838, \dodoi{10.1126/science.1256758}

\bibitem[{Tsai {et~al.}(2021b)Tsai, Innes, Lichtenberg, Taylor, Malik, Chubb,
  \& Pierrehumbert}]{Tsai2021b}
Tsai, S.-M., Innes, H., Lichtenberg, T., {et~al.} 2021b, The Astrophysical
  Journal Letters, 922, L27, \dodoi{10.3847/2041-8213/ac399a}

\bibitem[{Tsai {et~al.}(2017)Tsai, Lyons, Grosheintz, Rimmer, Kitzmann, \&
  Heng}]{tsai17}
Tsai, S.-M., Lyons, J.~R., Grosheintz, L., {et~al.} 2017, Astrophys. J. Suppl.
  Ser., 228, 1, \dodoi{10.3847/1538-4365/228/2/20}

\bibitem[{Tsai {et~al.}(2021)Tsai, Malik, Kitzmann, Lyons, Fateev, Lee, \&
  Heng}]{Tsai2021}
Tsai, S.-M., Malik, M., Kitzmann, D., {et~al.} 2021, The Astrophysical Journal,
  923, 264, \dodoi{10.3847/1538-4357/ac29bc}

\bibitem[{{Tsai} {et~al.}(2023){Tsai}, {Steinrueck}, {Parmentier}, {Lewis}, \&
  {Pierrehumbert}}]{Tsai2023}
{Tsai}, S.-M., {Steinrueck}, M., {Parmentier}, V., {Lewis}, N., \&
  {Pierrehumbert}, R. 2023, \mnras, 520, 3867, \dodoi{10.1093/mnras/stad214}

\bibitem[{Tsai {et~al.}(2023b)Tsai, Lee, Powell, Gao, Zhang, Moses,
  H{\'e}brard, Venot, Parmentier, Jordan, Hu, Alam, Alderson, Batalha, Bean,
  Benneke, Bierson, Brady, Carone, Carter, Chubb, Inglis, Leconte, Line,
  L{\'o}pez-Morales, Miguel, Molaverdikhani, Rustamkulov, Sing, Stevenson,
  Wakeford, Yang, Aggarwal, Baeyens, Barat, de~Val-Borro, Daylan, Fortney,
  France, Goyal, Grant, Kirk, Kreidberg, Louca, Moran, Mukherjee, Nasedkin,
  Ohno, Rackham, Redfield, Taylor, Tremblin, Visscher, Wallack, Welbanks,
  Youngblood, Ahrer, Batalha, Behr, Berta-Thompson, Blecic, Casewell,
  Crossfield, Crouzet, Cubillos, Decin, D{\'e}sert, Feinstein, Gibson,
  Harrington, Heng, Henning, Kempton, Krick, Lagage, Lendl, Lothringer,
  Mansfield, Mayne, Mikal-Evans, Palle, Schlawin, Shorttle, Wheatley, \&
  Yurchenko}]{Tsai2023b}
Tsai, S.-M., Lee, E. K.~H., Powell, D., {et~al.} 2023b, Nature,
  \dodoi{10.1038/s41586-023-05902-2}

\bibitem[{{Turrini} {et~al.}(2021){Turrini}, {Schisano}, {Fonte}, {Molinari},
  {Politi}, {Fedele}, {Pani{\'c}}, {Kama}, {Changeat}, \&
  {Tinetti}}]{Turrini2021}
{Turrini}, D., {Schisano}, E., {Fonte}, S., {et~al.} 2021, \apj, 909, 40,
  \dodoi{10.3847/1538-4357/abd6e5}

\bibitem[{van~der Walt {et~al.}(2011)van~der Walt, Colbert, \&
  Varoquaux}]{numpy}
van~der Walt, S., Colbert, S.~C., \& Varoquaux, G. 2011, Computing in Science
  Engineering, 13, 22, \dodoi{10.1109/MCSE.2011.37}

\bibitem[{{Venot} {et~al.}(2020b){Venot}, {Cavali{\'e}}, {Bounaceur},
  {Tremblin}, {Brouillard}, \& {Lhoussaine Ben Brahim}}]{Venot2020b}
{Venot}, O., {Cavali{\'e}}, T., {Bounaceur}, R., {et~al.} 2020b, \aap, 634,
  A78, \dodoi{10.1051/0004-6361/201936697}

\bibitem[{{Venot} {et~al.}(2020){Venot}, {Parmentier}, {Blecic}, {Cubillos},
  {Waldmann}, {Changeat}, {Moses}, {Tremblin}, {Crouzet}, {Gao}, {Powell},
  {Lagage}, {Dobbs-Dixon}, {Steinrueck}, {Kreidberg}, {Batalha}, {Bean},
  {Stevenson}, {Casewell}, \& {Carone}}]{Venot2020}
{Venot}, O., {Parmentier}, V., {Blecic}, J., {et~al.} 2020, \apj, 890, 176,
  \dodoi{10.3847/1538-4357/ab6a94}

\bibitem[{{Vuitton} {et~al.}(2021){Vuitton}, {Moran}, {He}, {Wolters},
  {Flandinet}, {Orthous-Daunay}, {Moses}, {Valenti}, {Lewis}, \&
  {H{\"o}rst}}]{Vuitton2021}
{Vuitton}, V., {Moran}, S.~E., {He}, C., {et~al.} 2021, \psj, 2, 2,
  \dodoi{10.3847/PSJ/abc558}

\bibitem[{{Wardenier} {et~al.}(2021){Wardenier}, {Parmentier}, {Lee}, {Line},
  \& {Gharib-Nezhad}}]{Joost2021}
{Wardenier}, J.~P., {Parmentier}, V., {Lee}, E. K.~H., {Line}, M.~R., \&
  {Gharib-Nezhad}, E. 2021, \mnras, 506, 1258, \dodoi{10.1093/mnras/stab1797}

\bibitem[{{Yates} {et~al.}(2020){Yates}, {Palmer}, {Manners}, {Boutle},
  {Kohary}, {Mayne}, \& {Abraham}}]{Yates2020}
{Yates}, J.~S., {Palmer}, P.~I., {Manners}, J., {et~al.} 2020, \mnras, 492,
  1691, \dodoi{10.1093/mnras/stz3520}

\bibitem[{Zahnle {et~al.}(2016)Zahnle, Marley, Morley, \& Moses}]{Zahnle2016}
Zahnle, K., Marley, M.~S., Morley, C.~V., \& Moses, J.~I. 2016, The
  Astrophysical Journal, 824, 137, \dodoi{10.3847/0004-637x/824/2/137}

\bibitem[{{Zamyatina} {et~al.}(2023){Zamyatina}, {H{\'e}brard}, {Drummond},
  {Mayne}, {Manners}, {Christie}, {Tremblin}, {Sing}, \&
  {Kohary}}]{Zamyatina2023}
{Zamyatina}, M., {H{\'e}brard}, E., {Drummond}, B., {et~al.} 2023, \mnras, 519,
  3129, \dodoi{10.1093/mnras/stac3432}

\bibitem[{{Zhang} {et~al.}(2019){Zhang}, {Chachan}, {Kempton}, \&
  {Knutson}}]{Zhang2019}
{Zhang}, M., {Chachan}, Y., {Kempton}, E. M.~R., \& {Knutson}, H.~A. 2019,
  \pasp, 131, 034501, \dodoi{10.1088/1538-3873/aaf5ad}

\bibitem[{{Zhang} {et~al.}(2020){Zhang}, {Chachan}, {Kempton}, {Knutson}, \&
  {Chang}}]{Zhang2020}
{Zhang}, M., {Chachan}, Y., {Kempton}, E. M.~R., {Knutson}, H.~A., \& {Chang},
  W.~H. 2020, \apj, 899, 27, \dodoi{10.3847/1538-4357/aba1e6}

\bibitem[{{Zhang} \& {Showman}(2018)}]{Xi2018}
{Zhang}, X., \& {Showman}, A.~P. 2018, \apj, 866, 2,
  \dodoi{10.3847/1538-4357/aada7c}

\end{thebibliography}


\end{document}